\documentclass[preprint,amssymb,amsmath,endfloats]{revtex4}

\pdfoutput=1
\usepackage{t1enc}
\usepackage[latin1]{inputenc}
\usepackage[english]{babel}
\usepackage[dvips]{color}
\usepackage{bm}
\usepackage{latexsym}
\usepackage{longtable}
\usepackage{theorem}
\usepackage{floatflt}
\usepackage{graphicx}
\begin{document}

\title{Photoproduction of $Z(4430)$ through mesonic Regge trajectories exchange.}

\author{Giuseppe Galat\`a}
\affiliation{Institut f\"ur Kernphysik, Johannes Gutenberg-Universit\"at, D-55099 Mainz, Germany}

\begin{abstract}
The recently discovered $Z(4430)$ mesonic resonance is believed to be a strong tetraquark candidate. The photoproduction in the channel $\gamma p\rightarrow Z^{+}(4430)n\rightarrow \psi ' \pi^{+} n$ has been proposed as the most effective way to confirm the $Z(4430)$ presence and to measure its quantum numbers.
In this work we present a model for high energy and forward angle $Z(4430)$ photoproduction in a effective Lagrangian approach. This model is based on the use of Regge trajectories exchange, thus a Regge propagator replaces the usual Feynman propagator. The differential and total cross sections and the asymmetries have been calculated for the $J^{P}$ quantum numbers $1^{-}$, $1^{+}$, $0^{-}$ and $2^{-}$ in the hypotheses that $Z(4430)$ has isospin $I=1$.
\end{abstract}

\maketitle

\section{\label{sec:introduzione} Introduction.}
The $Z(4430)$ meson has been observed by BELLE \cite{:2007wga} as a resonant structure in the $B\rightarrow K \psi ' \pi^{+} $ decay with a mass of $4433\pm 4\pm 2\;MeV$ and a width of $45^{+18+30}_{-13-13}\;MeV$. After this resonance has not been seen in the same channel by the BABAR collaboration \cite{:2008nk}, the BELLE collaboration performed a reanalysis of their data and confirmed the presence of the $Z^{+}(4430)$ resonance with new parameters $M_{Z}=4433^{+15+19}_{-12-13}\;MeV$ and $\Gamma _{Z}=107^{+87+74}_{-43-56}\;MeV$ \cite{:2009da}. Since the $Z$ decays in $\psi ' \pi^{+} $, it must have a positive charge and thus it cannot be a charmonium state but a state with isospin $I\geq 1$. In this work we hypothesize that the $Z(4430)$ is an isovector particle. Various different interpretations, then, have been proposed: a radially excited $c\bar cu \bar d$ tetraquark, a less compact hadronic meson molecule or be just due to threshold effects (see Ref. \cite{Branz:2010sh} and references therein).
However its strong coupling to $\psi ' \pi $ combined with its isospin $I=1$ makes it a strong candidate for a tetraquark resonance, as proposed long ago by Close and Lipkin \cite{Close:1978be}.

In order to seek to confirm in other experiments the observation of $Z(4430)$, the theoretical study of its production is an interesting topic. The production in nucleon-antinucleon scattering has been analized by Ke and Liu \cite{Ke:2008kf}, while a photoproduction search has been proposed by Liu, Zhao and Close \cite{Liu:2008qx} as the most effective. In this work we consider the $\gamma p\rightarrow Z^{+}(4430)n\rightarrow \psi ' \pi^{+} n$ meson photoproduction channel following the model based on the exchange of dominant meson Regge trajectories in the $t$-channel already used to successfully describe the pion and kaon photoproduction in Ref. \cite{Guidal:1997hy}. The model presented here should describe well the photoproduction of $Z$ at high and intermediate energies, which correspond to laboratory energies larger than $5\;GeV$ \cite{Eden:1971jm}, and forward angles, because for larger momentum transfer the Regge trajectories behavior is not anymore linear. 

The $J^{P}$ quantum numbers of the $Z(4430)$ have not been measured yet, however from the decay $ Z^{+}(4430)\rightarrow \psi ' \pi^{+}$ we can deduce that the possible quantum numbers are $J^{P}=1^{+}$ in S wave and $0^{-},1^{-},2^{-}$ in P wave. To each of these quantum numbers corresponds a different decay model and thus different resulting cross sections. In section \ref{sec:modellilagrangiane} we calculate the differential and total cross sections for each $J^{P}$ quantum number of the $Z$ and for each exchanged Regge trajectories used, i.e. the $\pi $, the $\rho $ and the $a_{0}$ trajectories. In section \ref{sec:asimmetrie} the photon, target and recoil asymmetries are calculated in each case.

\section{\label{sec:formalismoRegge}The Regge formalism.}
In the fifties Regge \cite{Regge:1959mz,Regge:1960zc} showed the importance to extend the angular momentum $J$ to the complex field. Here a very brief and partial summary of the Regge theory is introduced, while for more general reviews see Refs. \cite{Corthals:2005ce,Storrow:1986zw,Eden:1971jm}.

Regge proved that bound states and resonances belong to families related by trajectory functions $\alpha _{R}(t)$. These functions give the position of the poles of the amplitude, which occur at $J=\alpha _{R}(t)$. For integral $J$ this equation gives the masses of the particles on the trajectory, i.e. $J=\alpha _{R}(m_{J}^{2})$.

The amplitudes of every generic particle exchange scattering $2\rightarrow 2$ may be written as a sum of Regge terms
\begin{equation}
T_{2\rightarrow 2}=\sum\limits_{R} T^{(R)}(s,t),
\end{equation}
where each term is given by a vertex function $\beta $ and a Reggeised propagator $\mathcal{P}_{Regge}$:
\begin{equation}
T^{(R)}(s,t)=\beta (t)\mathcal{P}_{Regge}(s,\alpha _{R}(t)).
\end{equation}
At small negative $t$ the Regge trajectories have been shown to have a linear form
\begin{equation}
\alpha _{R}(t)=\alpha _{R,0}+\alpha '_{R}(t-m^{2}_{R}).
\end{equation}
In this equation $m_{R}$ is the mass of the trajectory's lightest particle, which we refer to as first realization.
The constant $\alpha _{R,0}$ is the spin $J$ of the first realization of the trajectory, since we know that in the amplitude poles, with integral $J$, $J=\alpha _{R}(m_{J}^{2})$. The slope of the trajectory $\alpha '_{R}$ varies only slightly among the various trajectories and can usually be considered a universal feature of the theory with a value around $0.8-0.9\;GeV^{-2}$. 

The Regge propagator is
\begin{equation}
\mathcal{P}_{Regge}=(\frac{s}{s_{0}})^{\alpha _{R} (t)-\alpha _{R,0}}\frac{\pi \alpha '_{R}}{\sin (\pi (\alpha _{R} (t)-\alpha _{R,0}))}\frac{\mathcal{S}_{R}+e^{-i\pi (\alpha _{R} (t)-\alpha _{R,0})}}{2\Gamma (1+\alpha _{R}(t)-\alpha _{R,0})},
\end{equation}  
where $s,t$ are the usual Mandelstam variables, $s_{0}=1\;GeV^{2}$ is a mass scale and $\mathcal{S}_{R }=\pm 1$ is the signature of the trajectory. It is important to note that at high $s$ and small $t$ the magnitude of the Regge propagator is given mainly by the $(\frac{s}{s_{0}})^{\alpha _{R} (t)-\alpha _{R,0}}$ factor. The exponent $\alpha _{R} (t)-\alpha _{R,0}=\alpha '_{R}(t-m^{2}_{R})$  implies, then, that trajectories with lighter first realization particles will tend to dominate. 

In most mesonic cases, the trajectories with signature $\mathcal{S}_{R}=1$ and $-1$ coincide. In this occurrence they are called degenerate.
Then the amplitudes for $\mathcal{S}_{R}=1$ and $-1$ can be added in order to obtain the new propagator
\begin{equation}
\mathcal{P}_{Regge}=(\frac{s}{s_{0}})^{\alpha _{R} (t)-\alpha _{R,0}}\frac{\pi \alpha '_{R}}{\sin (\pi (\alpha _{R} (t)-\alpha _{R,0}))}\frac{e^{-i\pi (\alpha _{R} (t)-\alpha _{R,0})}}{\Gamma (1+\alpha _{R} (t)-\alpha _{R,0})} 
\end{equation} 
or subtracted to obtain
\begin{equation}
\mathcal{P}_{Regge}=(\frac{s}{s_{0}})^{\alpha _{R} (t)-\alpha _{R,0}}\frac{\pi \alpha '_{R}}{\sin (\pi (\alpha _{R} (t)-\alpha _{R,0}))}\frac{1}{\Gamma (1+\alpha _{R} (t)-\alpha _{R,0})}.
\end{equation}
The difference between these two propagators are only in a phase factor, thus it does not matter for the calculation of the cross section for a single exchanged trajectory but only in the interference terms between different trajectories. Whether to use a degenerate propagator or not is something that depends on the process considered. In fact non degenerate trajectories produce dips in the differential cross section, called wrong-signature zeros, which correspond to poles of the gamma function not removed by the sine function, while degenerate trajectories produce smooth curves. 

In this work we opt to embed the Regge formalism into a tree level
effective-Lagrangian model, following the approach developed by Levy, Majerotto and Read \cite{Levy:1973aq} and Guidal, Laget and
Vanderhaeghen \cite{Guidal:1997hy} in their treatment of high-energy $\pi $ and $K$ photoproduction . In this approach the exchange of a Regge trajectory is taken into account by replacing the usual Feynman propagator of a single particle with the Regge propagator, while keeping the vertex structure given by the Feynman diagrams which correspond to the first realization of the trajectory. This approximate method to implement the Regge formalism is the only one possible for the $Z(4430)$ photoproduction because the known data are not enough to determine the vertex function $\beta (t)$ through a fit.

\section{\label{sec:modellilagrangiane}$Z(4430)$ decay models.}
We have seen that the importance in determining the cross section decrease exponentially with the increase of the mass of the first realization of the exchanged trajectory.  
For this reason we can limit to consider only the three main contributions to the $Z(4430)$ photoproduction on nucleons through meson exchange: the $\pi $ exchange, which is by far the principal one, and the $\rho $ and the $a_{0}(980)$ exchanges, whose contributions to the total cross section are negligible but are nevertheless important for the calculation of the polarization asymmetries. Moreover, when the degenerate Regge propagator is used, the contributions of the corresponding degenerate trajectories, i.e. respectively the $b_{1}(1235)$, $a_{2}(1320)$ and $\rho (1700)$ ones, are automatically taken into account. 

The coupling of $Z$ to the photon is derived through the VMD (Vector Meson Dominance) model by assuming that the coupling is due to a sum of intermediate vector mesons which connect the photon with the $Z$ and the exchanged meson. The $\psi '$ is the dominant intermediate vector meson and its coupling to the photon is given by 
\begin{equation}
\mathcal{L}_{\psi '\gamma }=\frac{e M^{2}_{\psi '}}{f_{\psi '}}\psi '_{\mu }A^{\mu },
\end{equation}
where the constant $\frac{e}{f_{\psi '}}=0.0166$ is determined by the decay $\psi '\rightarrow e^{+}e^{-}$ \cite{Klingl:1996by,Yao:2006px}.

Second-order terms, like the $J/\psi $ intermediate vector meson contribution, are evaluated in Ref. \cite{Liu:2008qx} and found to be probably very small.
\subsection{\label{subsec:piexchange}The $\pi $ exchange contribution.} 
We will first consider the process $\gamma p\rightarrow Z^{+}n$ with the exchange of a pionic Regge trajectory. 
\begin{figure}[h]
\includegraphics[width=8cm]{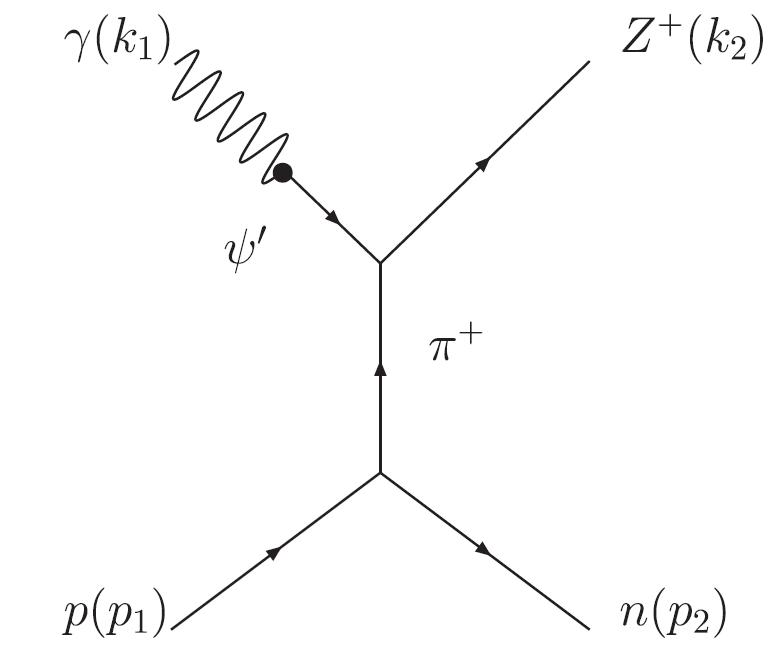}  
\caption{\label{fig:Zpiexchange}Graphic representation of the $\gamma p\rightarrow Z^{+}n$ process with a pion exchange.}
\end{figure}

The pion-nucleon coupling is described by the Lagrangian
\begin{equation}
\mathcal{L}_{\pi NN}=-i\sqrt{2}g_{\pi NN}\bar n \gamma _{5} p \pi ^{-},
\end{equation}
where $g^{2}_{\pi NN}=14\cdot 4\pi $. The following form factor is also applied to the vertex:
\begin{equation}
F_{\pi NN}=\frac{\Lambda ^{2}_{\pi }-m^{2}_{\pi }}{\Lambda ^{2}_{\pi }-q^{2}},
\end{equation}
where $\Lambda _{\pi }=0.7\;GeV$ and $q^{2}$ is the squared transferred momentum.

\subsubsection{\label{subsubsec:1-}The $J^{P}=1^{-}$ case.}
In the case that $Z(4430)$ has $J^{P}=1^{-}$ the $Z\psi '\pi $ coupling is given by \cite{Liu:2008qx}
\begin{equation}
\mathcal{L}_{Z\psi '\pi }=\frac{g_{Z\psi '\pi }}{M_{Z}}\epsilon _{\mu \nu \alpha \beta }\partial ^{\mu }\psi '^{\nu }\partial ^{\alpha }Z^{+\beta }\pi ^{-}+h.c.
\end{equation}
The coupling constant $g_{Z\psi '\pi }/M_{Z}=2.365\;GeV^{-1}$ is determined by the decay $Z\rightarrow \psi '\pi $. Again a form factor is applied to the vertex:
\begin{equation}
F_{Z\psi '\pi }=\frac{M^{2}_{\psi '}-m^{2}_{\pi }}{M^{2}_{\psi '}-q^{2}}.
\end{equation}

The resulting amplitude for the $\gamma p\rightarrow Z^{+}(4430)n$ process via pion exchange is 
\begin{equation}
T_{1-}=-i(\sqrt{2}g_{\pi NN}\frac{g_{Z\psi '\pi }}{M_{Z}}\frac{e}{f_{\psi '}})\bar u(p_{2})\gamma _{5}u(p_{1})\epsilon _{\mu \nu \alpha \beta }k_{1}^{\mu }k_{2}^{\alpha }\varepsilon ^{\nu }_{\gamma }(k_{1})\varepsilon ^{*\beta }_{Z}(k_{2})\frac{1}{q^{2}-m_{\pi }^{2}}F_{\pi NN}F_{Z\psi '\pi }.
\end{equation}
The four vector momenta are attributed to each particle as in Fig. \ref{fig:Zpiexchange}.

In the Regge trajectories approach the Feynman propagator $\frac{1}{q^{2}-m_{\pi }^{2}}$ is replaced by the Regge propagator, which in case of pion trajectory is
\begin{equation}
\mathcal{P}^{\pi }_{Regge}=(\frac{s}{s_{0}})^{\alpha _{\pi} (t)}\frac{\pi \alpha '_{\pi }}{\sin (\pi \alpha _{\pi} (t))}\frac{\mathcal{S}_{\pi }+e^{-i\pi\alpha _{\pi} (t)}}{2\Gamma (1+\alpha _{\pi }(t))}, 
\end{equation}
where the signature $\mathcal{S}_{\pi }=1$ and the pion Regge trajectory $\alpha _{\pi} (t)$ is defined as \cite{Guidal:1997hy}
\begin{equation}
\alpha _{\pi} (t)=\alpha _{\pi 0}+\alpha '_{\pi }t=0.7(t-m_{\pi }^{2}).
\end{equation}

%We can now calculate both the differential cross section $\frac{d\sigma }{dt}$ and the total cross section $\sigma $. The results can be found respectively in Fig. \ref{fig:Zpi1-Diff} and \ref{fig:Zpi1-Tot}.
%\begin{figure}
%\includegraphics[width=10cm]{Zpi1-Diff.jpg}  
%\caption{\label{fig:Zpi1-Diff}Differential cross section for the photoproduction of $Z^{+}(4430)$ with $J^{P}=1^{-}$ using a non degenerate pionic Regge trajectory.}
%\end{figure}

%\begin{figure}
%\includegraphics[width=10cm]{Zpi1-Tot.jpg}  
%\caption{\label{fig:Zpi1-Tot}Total cross section for the photoproduction of $Z^{+}(4430)$ with $J^{P}=1^{-}$ using a non degenerate pionic Regge trajectory.}
%\end{figure}

\begin{figure}
\includegraphics[width=10cm]{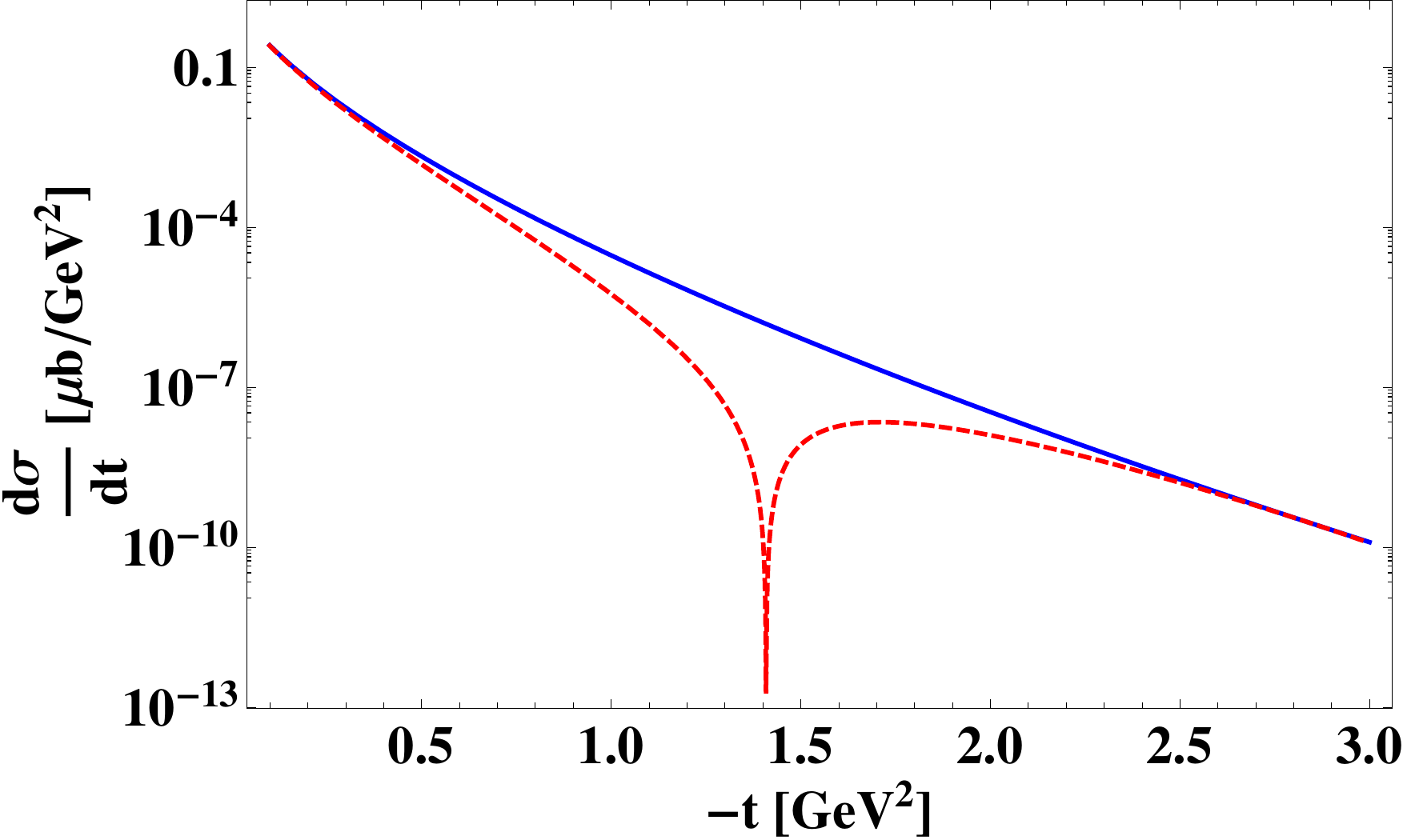}  
\caption{\label{fig:Zpi1-DegNonDegDiff}Differential cross section for the photoproduction of $Z^{+}(4430)$ with $J^{P}=1^{-}$ using a degenerate (blue solid line) and a non degenerate (red dashed line) pionic Regge trajectory.}
\end{figure}

\begin{figure}
\includegraphics[width=10cm]{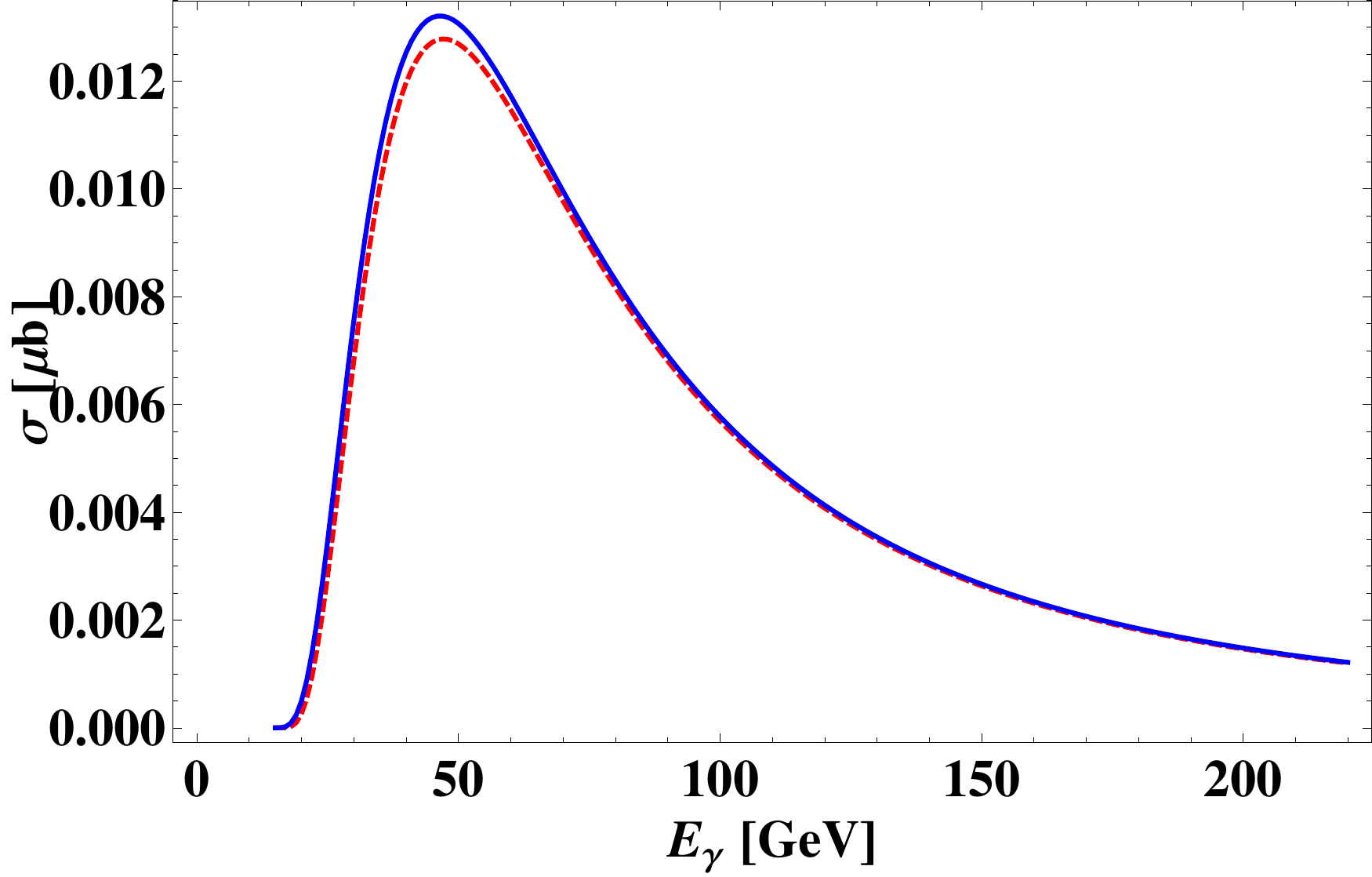}  
\caption{\label{fig:Zpi1-DegNonDegTot}Total cross section for the photoproduction of $Z^{+}(4430)$ with $J^{P}=1^{-}$ using a degenerate (blue solid line) and a non degenerate (red dashed line) pionic Regge trajectory.}
\end{figure}

If the degenerate trajectory is also considered, the amplitudes for $\mathcal{S}=1$ and $-1$ can be added in order to obtain the new propagator
\begin{equation}
\label{eq:propdegadded}
\mathcal{P}^{\pi }_{Regge}=(\frac{s}{s_{0}})^{\alpha _{\pi} (t)}\frac{\pi \alpha '_{\pi }}{\sin (\pi \alpha _{\pi} (t))}\frac{e^{-i\pi \alpha_{\pi } (t)}}{\Gamma (1+\alpha _{\pi }(t))} 
\end{equation} 
or subtracted to obtain
\begin{equation}
\label{eq:propdegsub}
\mathcal{P}^{\pi }_{Regge}=(\frac{s}{s_{0}})^{\alpha _{\pi} (t)}\frac{\pi \alpha '_{\pi }}{\sin (\pi \alpha _{\pi} (t))}\frac{1}{\Gamma (1+\alpha _{\pi }(t))}.
\end{equation}
The results for the differential cross section in the degenerate and non degenerate case are shown in Fig. \ref{fig:Zpi1-DegNonDegDiff}, while the total cross section for both cases can be found in Fig. \ref{fig:Zpi1-DegNonDegTot}. As we can see in these figures, the non degenerate trajectory produces the expected dip in the differential cross section but the difference in the total cross section is negligible.

In analogy with the pion photoproduction \cite{Guidal:1997hy,Sibirtsev:2007wk,Chew:1957tf} the added degenerate propagator, Eq. \ref{eq:propdegadded}, is used for the $Z^{+}$ photoproduction and the subtracted one, Eq. \ref{eq:propdegsub}, for the $Z^{-}$.

\subsubsection{\label{subsubsec:1+}The $J^{P}=1^{+}$ case.}
The difference from the previous case lies in the new $Z\psi '\pi $ coupling given by \cite{Liu:2008qx}
\begin{equation}
\mathcal{L}_{Z\psi '\pi }=\frac{g_{Z\psi '\pi }}{M_{Z}}(\partial ^{\alpha }\psi '^{\beta }\partial _{\alpha }\pi Z_{\beta }-\partial ^{\alpha }\psi '^{\beta }\partial _{\beta }\pi Z_{\alpha }),
\end{equation}
with $g_{Z\psi '\pi }/M_{Z}=2.01\;GeV^{-1}$. 

%Again we show the differential and total cross section in both cases of trajectory degeneration.

%\begin{figure}
%\includegraphics[width=10cm]{Zpi1+Diff.jpg}  
%\caption{\label{fig:Zpi1+Diff}Differential cross section for the photoproduction of $Z^{+}(4430)$ with $J^{P}=1^{+}$ using a non degenerate pionic Regge trajectory.}
%\end{figure}

%\begin{figure}
%\includegraphics[width=10cm]{Zpi1+Tot.jpg}  
%\caption{\label{fig:Zpi1+Tot}Total cross section for the photoproduction of $Z^{+}(4430)$ with $J^{P}=1^{+}$ using a non degenerate pionic Regge trajectory.}
%\end{figure}

%\begin{figure}
%\includegraphics[width=10cm]{Zpi1+DegDiff.jpg}  
%\caption{\label{fig:Zpi1+DegDiff}Differential cross section for the photoproduction of $Z^{+}(4430)$ with $J^{P}=1^{+}$ using a degenerate pionic Regge trajectory.}
%\end{figure}

%\begin{figure}
%\includegraphics[width=10cm]{Zpi1+DegTot.jpg}  
%\caption{\label{fig:Zpi1+DegTot}Total cross section for the photoproduction of $Z^{+}(4430)$ with $J^{P}=1^{+}$ using a  degenerate pionic Regge trajectory.}
%\end{figure}

\subsubsection{\label{subsubsec:0-}The $J^{P}=0^{-}$ case.}
In this case the $Z\psi '\pi $ coupling is \cite{Liu:2008qx}
\begin{equation}
\mathcal{L}_{Z\psi '\pi }=i\frac{g_{Z\psi '\pi }}{M_{Z}}(\pi ^{-}\partial _{\mu }Z^{+}-\partial _{\mu }\pi ^{-}Z^{+})\psi '^{\mu },
\end{equation}
with $g_{Z\psi '\pi }/M_{Z}=1.39\;GeV^{-1}$. It is important to notice that the quantum numbers for the $Z^{0}$ in this case would be $J^{PC}=0^{--}$, which is an explicitly exotic combination.

%The differential and total cross section are in Figs. \ref{fig:Zpi0-Diff} and \ref{fig:Zpi0-Tot} for the non degenerate case and Figs. \ref{fig:Zpi0-DegDiff} and \ref{fig:Zpi0-DegDiff} for the degenerate one. 

%\begin{figure}
%\includegraphics[width=10cm]{Zpi0-Diff.jpg}  
%\caption{\label{fig:Zpi0-Diff}Differential cross section for the photoproduction of $Z^{+}(4430)$ with $J^{P}=0^{-}$ using a non degenerate pionic Regge trajectory.}
%\end{figure}

%\begin{figure}
%\includegraphics[width=10cm]{Zpi0-Tot.jpg}  
%\caption{\label{fig:Zpi0-Tot}Total cross section for the photoproduction of $Z^{+}(4430)$ with $J^{P}=0^{-}$ using a non degenerate pionic Regge trajectory.}
%\end{figure}

%\begin{figure}
%\includegraphics[width=10cm]{Zpi0-DegDiff.jpg}  
%\caption{\label{fig:Zpi0-DegDiff}Differential cross section for the photoproduction of $Z^{+}(4430)$ with $J^{P}=0^{-}$ using a degenerate pionic Regge trajectory.}
%\end{figure}

%\begin{figure}
%\includegraphics[width=10cm]{Zpi0-DegTot.jpg}  
%\caption{\label{fig:Zpi0-DegTot}Total cross section for the photoproduction of $Z^{+}(4430)$ with $J^{P}=0^{-}$ using a  degenerate pionic Regge trajectory.}
%\end{figure}

\subsubsection{\label{subsubsec:2-}The $J^{P}=2^{-}$ case.}
In the tensor case the $Z\psi '\pi $ coupling is \cite{Ke:2008kf}
\begin{equation}
\mathcal{L}_{Z\psi '\pi }=ig_{Z\psi '\pi }\psi '_{\mu }Z^{\mu \nu }\partial _{\nu }\pi ^{-},
\end{equation}
with $g_{Z\psi '\pi }=0.32$. 

%The differential and total cross section are in Figs. \ref{fig:Zpi2-Diff} and \ref{fig:Zpi2-Tot} for the non degenerate case and Figs. \ref{fig:Zpi2-DegDiff} and \ref{fig:Zpi2-DegTot} for the degenerate one.

%\begin{figure}
%\includegraphics[width=10cm]{Zpi2-Diff.jpg}  
%\caption{\label{fig:Zpi2-Diff}Differential cross section for the photoproduction of $Z^{+}(4430)$ with $J^{P}=2^{-}$ using a non degenerate pionic Regge trajectory.}
%\end{figure}

%\begin{figure}
%\includegraphics[width=10cm]{Zpi2-Tot.jpg}  
%\caption{\label{fig:Zpi2-Tot}Total cross section for the photoproduction of $Z^{+}(4430)$ with $J^{P}=2^{-}$ using a non degenerate pionic Regge trajectory.}
%\end{figure}

%\begin{figure}
%\includegraphics[width=10cm]{Zpi2-DegDiff.jpg}  
%\caption{\label{fig:Zpi2-DegDiff}Differential cross section for the photoproduction of $Z^{+}(4430)$ with $J^{P}=2^{-}$ using a degenerate pionic Regge trajectory.}
%\end{figure}

%\begin{figure}
%\includegraphics[width=10cm]{Zpi2-DegTot.jpg}  
%\caption{\label{fig:Zpi2-DegTot}Total cross section for the photoproduction of $Z^{+}(4430)$ with $J^{P}=2^{-}$ using a  degenerate pionic Regge trajectory.}
%\end{figure}

\begin{figure}
\includegraphics[width=10cm]{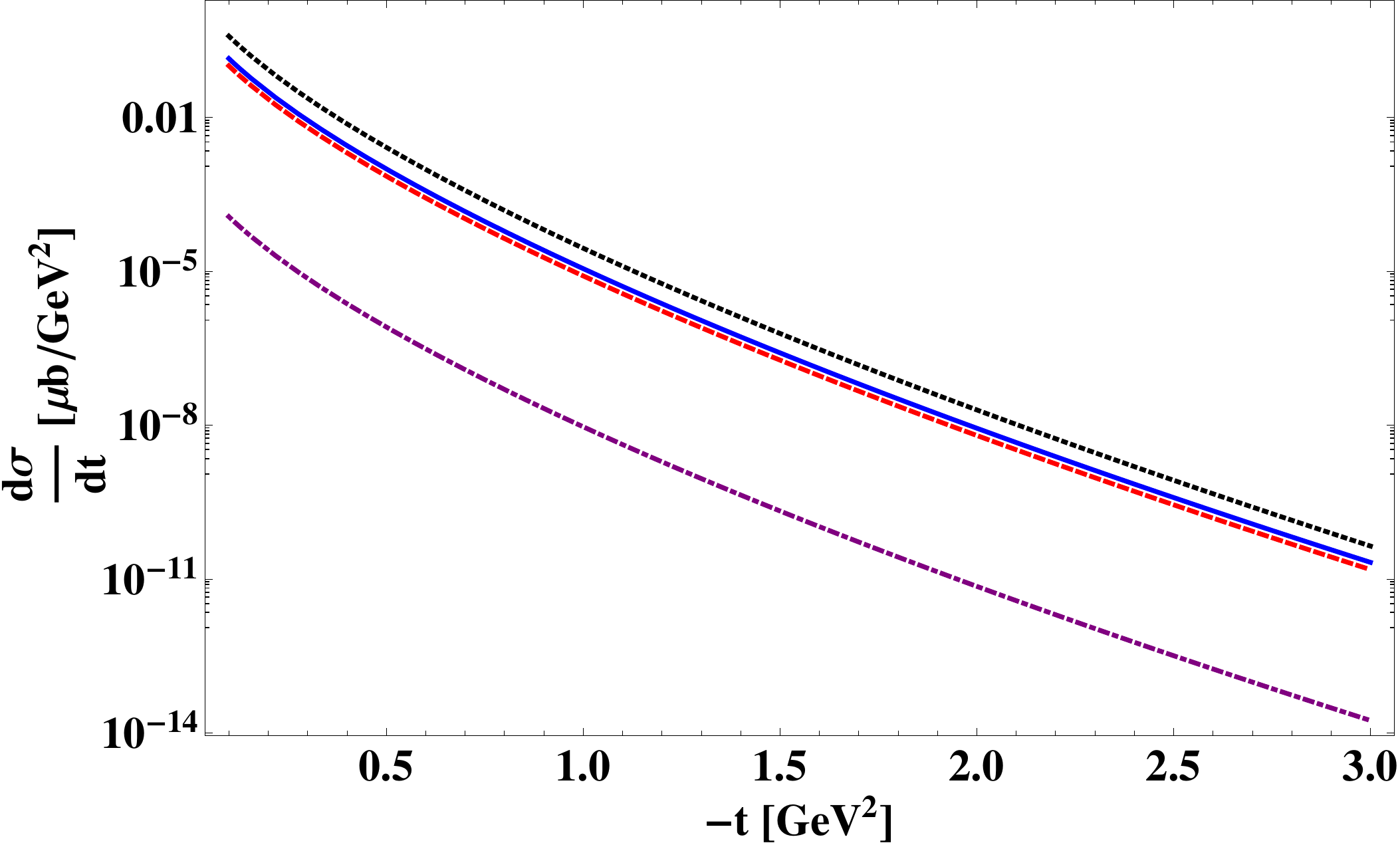}  
\caption{\label{fig:ZDiffCrosssecspi}Differential cross section for the photoproduction (photon energy $30\;GeV$) of $Z^{+}(4430)$ with $J^{P}=1^{-}$ (Blue solid line), $J^{P}=1^{+}$ (Red dashed line), $J^{P}=0^{-}$ (Black dotted line) and $J^{P}=2^{-}$ (Purple dot-dashed line) using a degenerate pionic Regge trajectory.}
\end{figure}

The differential cross-sections for all the four possible quantum numbers examined ($1^{-},1^{+},0^{-},2^{-}$) in the degenerate trajectories case are shown in Figure \ref{fig:ZDiffCrosssecspi}. 

\begin{figure}
\includegraphics[width=10cm]{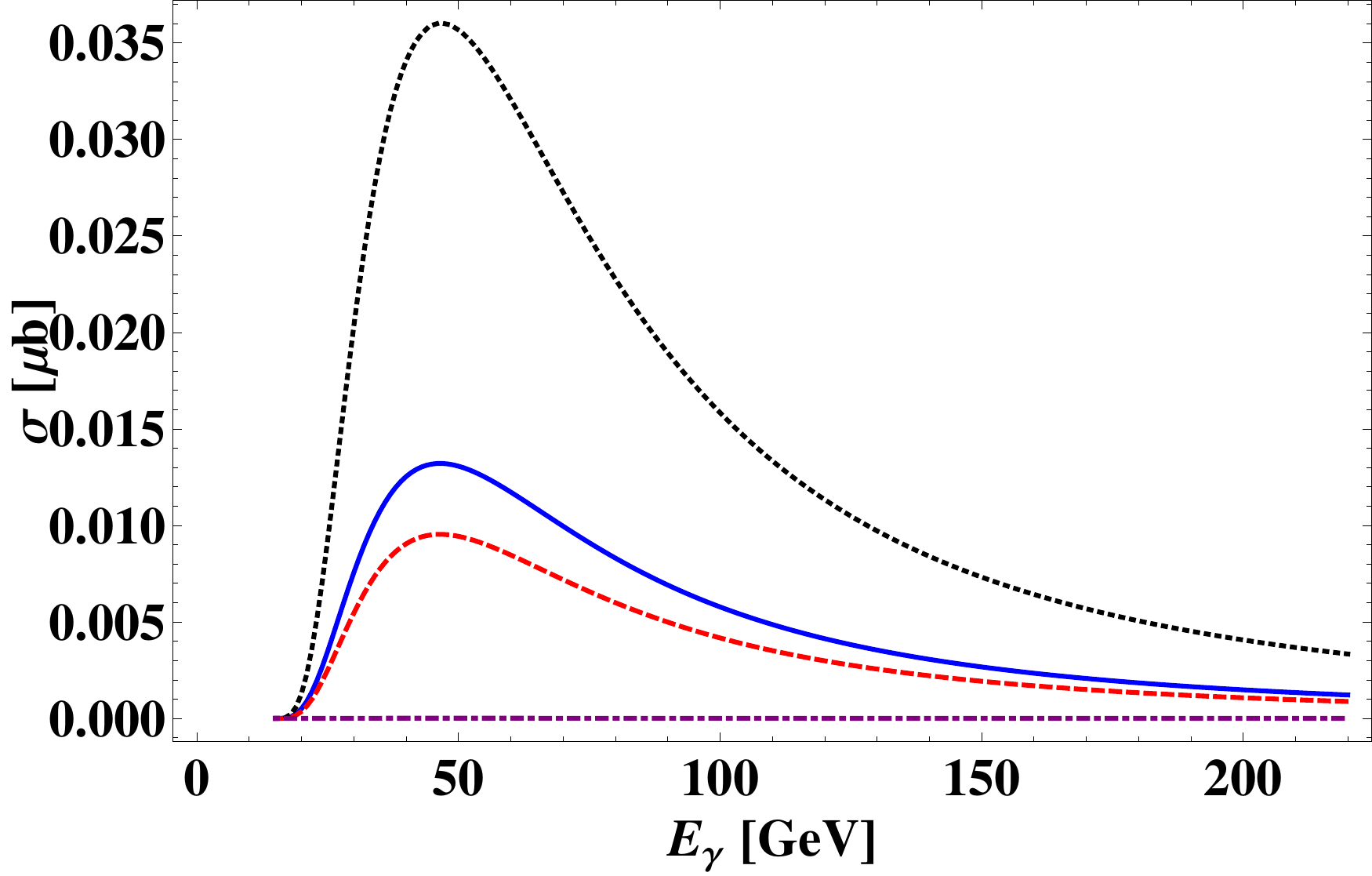}  
\caption{\label{fig:ZTotCrosssecspilin}Total cross section for the photoproduction of $Z^{+}(4430)$ with $J^{P}=1^{-}$ (Blue solid line), $J^{P}=1^{+}$ (Red dashed line), $J^{P}=0^{-}$ (Black dotted line) and $J^{P}=2^{-}$ (Purple dot-dashed line) using a degenerate pionic Regge trajectory.}
\end{figure}

\begin{figure}
\includegraphics[width=10cm]{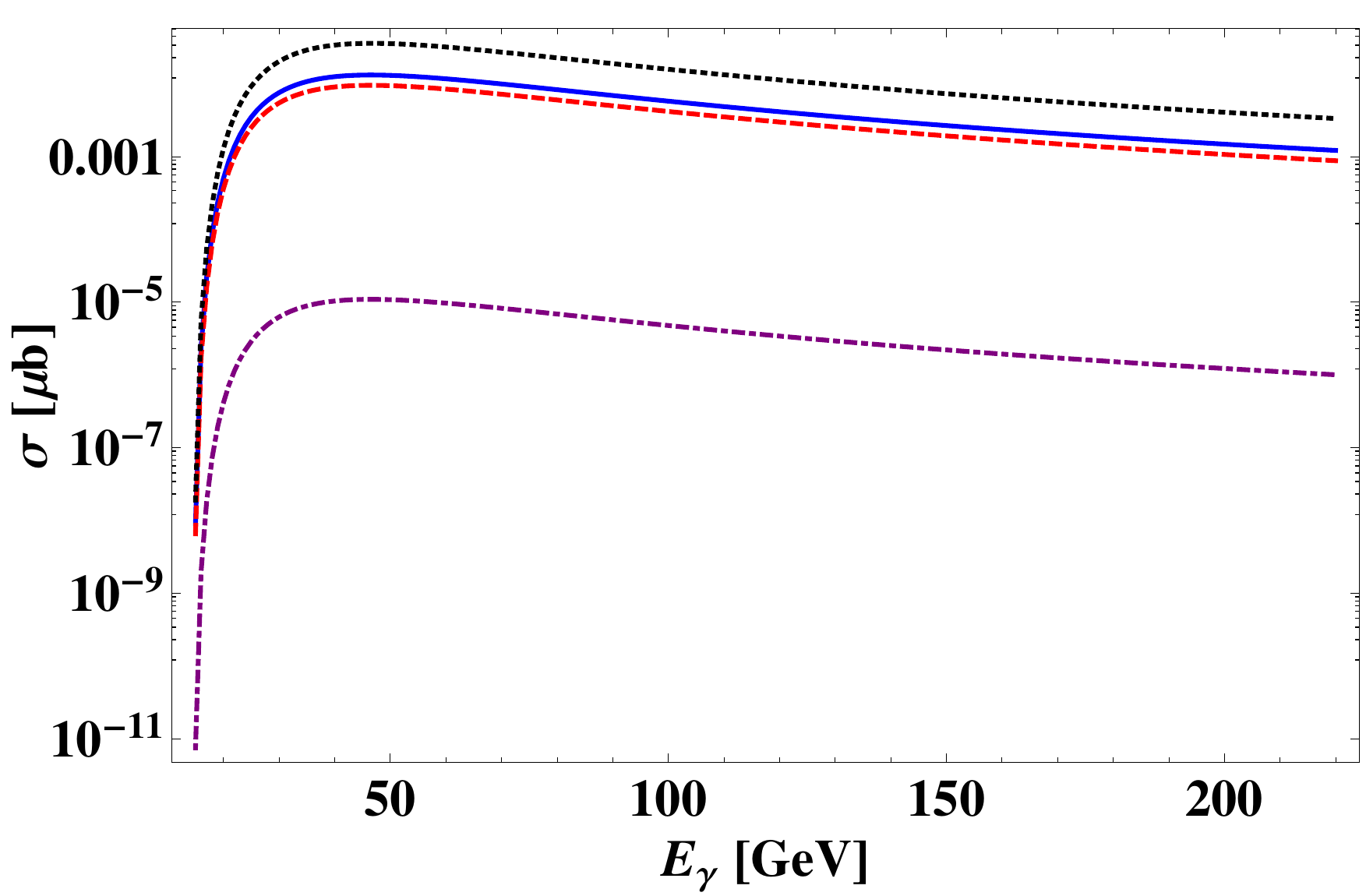}  
\caption{\label{fig:ZTotCrosssecspilog}Same as Fig. \ref{fig:ZTotCrosssecspilin} but with a logarithmic scale.}
\end{figure}

The total cross sections are in Figures \ref{fig:ZTotCrosssecspilin} and \ref{fig:ZTotCrosssecspilog}. From these figures it is evident that the $J^{P}=2^{-}$ cross section is several orders of magnitude smaller than the other cross sections and would be completely covered by the background (for an analysis of the background see Ref. \cite{Liu:2008qx}). 

\subsection{\label{subsec:a0}The $a_{0}(980)$ exchange contribution.}
The same $\gamma p\rightarrow Z^{+}n$ process can also happen through the exchange of the $a_{0}(980)$ meson (and its trajectory in the Regge approach). The possible $Z$ quantum numbers in this case would be $J^{P}=1^{-}$ in S wave and $0^{+},1^{+},2^{+}$ in P wave. Since the $a_{0}(980)$ amplitude is expected to be very small, we are interested to calculate only the cases where the $a_{0}$ can interfere with the pionic amplitude, i.e. $J^{P}=1^{+},1^{-}$. 

The Lagrangian for the $a_{0}NN$ vertex is \cite{Liu:2008qx}
\begin{equation}
\mathcal{L}_{a_{0} NN}=\sqrt{2}g_{a_{0} NN}\bar n p a_{0},
\end{equation}
where $g^{2}_{a_{0} NN}/4\pi=1.075$. The form factor applied to the vertex is
\begin{equation}
F_{a_{0} NN}=\frac{\Lambda ^{2}_{a_{0}}-m^{2}_{a_{0}}}{\Lambda ^{2}_{a_{0}}-q^{2}},
\end{equation}
with $\Lambda _{a_{0}}=2.0\;GeV$.

\subsubsection{\label{subsubsec:a01-}The $J^{P}=1^{-}$ case.}
In the $J^{P}=1^{-}$ case, the $Z\psi 'a_{0} $ coupling Lagrangian is \cite{Liu:2008qx}
\begin{equation}
\mathcal{L}_{Z\psi 'a_{0} }=\frac{g_{Z\psi 'a_{0} }}{M_{Z}}(\partial ^{\alpha }Z^{\beta }\partial _{\alpha }\psi '_{\beta }-\partial ^{\alpha }Z^{\beta }\partial _{\beta }\psi '_{\alpha })a_{0}.
\end{equation}
Since there are no data on the coupling constant $g_{Z\psi 'a_{0} }$, it is set equal to $g_{Z\psi '\pi }$ as an upper limit estimation.

In substituting the Feynman propagator with the Regge one we follow the same procedure of the pionic exchange, with the only difference that we have to replace the pion trajectory with the $a_{0}$ one. The Regge trajectory for the $a_{0}$ is not well established but we chose the one found in Ref. \cite{Gerasyuta:1998fc}:
\begin{equation}
\alpha _{a_{0}}(t)=-0.5+0.6t.
\end{equation}
%The resulting differential and total cross sections are in Figs. \ref{fig:Za01-Diff}, \ref{fig:Za01-Tot}, \ref{fig:Za01-DegDiff} and \ref{fig:Za01-DegTot}. 

%\begin{figure}
%\includegraphics[width=10cm]{Za01-Diff.jpg}  
%\caption{\label{fig:Za01-Diff}Differential cross section for the photoproduction of $Z^{+}(4430)$ with $J^{P}=1^{-}$ using a non degenerate $a_{0}$ Regge trajectory.}
%\end{figure}

%\begin{figure}
%\includegraphics[width=10cm]{Za01-Tot.jpg}  
%\caption{\label{fig:Za01-Tot}Total cross section for the photoproduction of $Z^{+}(4430)$ with $J^{P}=1^{-}$ using a non degenerate pionic Regge trajectory.}
%\end{figure}

%\begin{figure}
%\includegraphics[width=10cm]{Za01-DegDiff.jpg}  
%\caption{\label{fig:Za01-DegDiff}Differential cross section for the photoproduction of $Z^{+}(4430)$ with $J^{P}=1^{-}$ using a degenerate $a_{0}$ Regge trajectory.}
%\end{figure}

%\begin{figure}
%\includegraphics[width=10cm]{Za01-DegTot.jpg}  
%\caption{\label{fig:Za01-DegTot}Total cross section for the photoproduction of $Z^{+}(4430)$ with $J^{P}=1^{-}$ using a  degenerate $a_{0}$ Regge trajectory.}
%\end{figure}
\subsubsection{\label{subsubsec:a01+}The $J^{P}=1^{+}$ case.}
If $Z(4430)$ has $J^{P}=1^{+}$ quantum numbers, the $Z\psi 'a_{0} $ vertex is described by \cite{Liu:2008qx}
\begin{equation}
\mathcal{L}_{Z\psi 'a_{0} }=\frac{g_{Z\psi 'a_{0} }}{M_{Z}}\epsilon _{\mu \nu \alpha \beta }\partial ^{\mu }\psi '^{\nu }\partial ^{\alpha }Z^{\beta }a_{0}.
\end{equation}
The coupling constant is again taken equal to the corresponding one of the pion exchange.

%In the Figs. \ref{fig:Za01+Diff}, \ref{fig:Za01+Tot}, \ref{fig:Za01+DegDiff} and \ref{fig:Za01+DegTot} there are the results for the differential and total cross section.
%\begin{figure}
%\includegraphics[width=10cm]{Za01+Diff.jpg}  
%\caption{\label{fig:Za01+Diff}Differential cross section for the photoproduction of $Z^{+}(4430)$ with $J^{P}=1^{+}$ using a non degenerate $a_{0}$ Regge trajectory.}
%\end{figure}

%\begin{figure}
%\includegraphics[width=10cm]{Za01+Tot.jpg}  
%\caption{\label{fig:Za01+Tot}Total cross section for the photoproduction of $Z^{+}(4430)$ with $J^{P}=1^{+}$ using a non degenerate pionic Regge trajectory.}
%\end{figure}

%\begin{figure}
%\includegraphics[width=10cm]{Za01+DegDiff.jpg}  
%\caption{\label{fig:Za01+DegDiff}Differential cross section for the photoproduction of $Z^{+}(4430)$ with $J^{P}=1^{+}$ using a degenerate $a_{0}$ Regge trajectory.}
%\end{figure}

%\begin{figure}
%\includegraphics[width=10cm]{Za01+DegTot.jpg}  
%\caption{\label{fig:Za01+DegTot}Total cross section for the photoproduction of $Z^{+}(4430)$ with $J^{P}=1^{+}$ using a  degenerate $a_{0}$ Regge trajectory.}
%\end{figure}

\begin{figure}
\includegraphics[width=10cm]{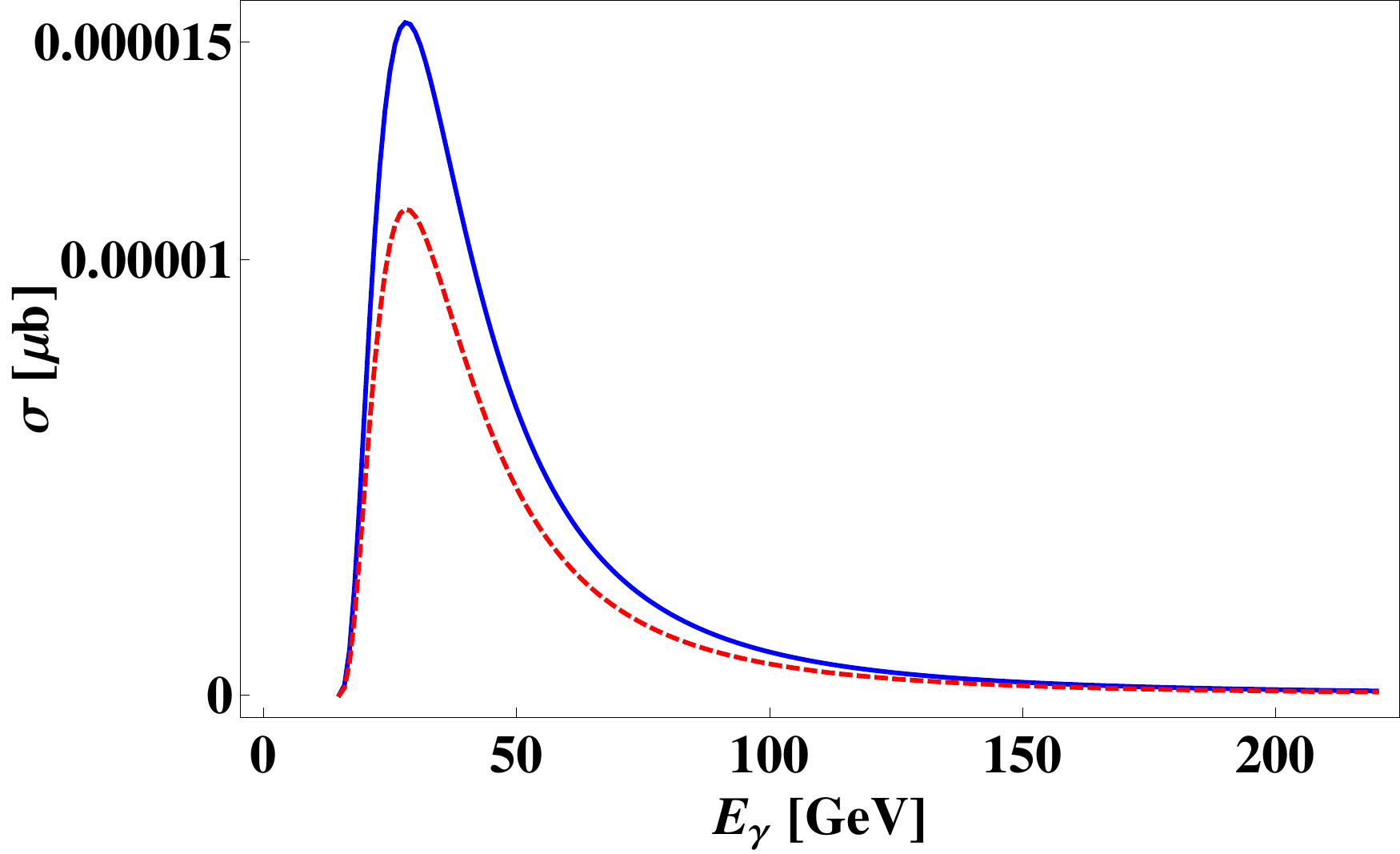}  
\caption{\label{fig:ZTotCrosssecsa0}Total cross section for the photoproduction of $Z^{+}(4430)$ with $J^{P}=1^{-}$ (Blue solid line) and $J^{P}=1^{+}$ (Red dashed line) using a degenerate $a_{0}$ Regge trajectory.}
\end{figure}

The total cross sections for both quantum numbers are given in Fig. \ref{fig:ZTotCrosssecsa0}.

\subsection{\label{subsec:rho}The $\rho (770)$ exchange case.}
The $\gamma p\rightarrow Z^{+}n$ process can also happen through the exchange of the $\rho$ meson (and its trajectory in the Regge approach). The possible $Z$ quantum numbers which interfere with the pion exchange channel in this case would be $J^{P}=1^{+}$ in S wave and $0^{-},1^{-},2^{-}$ in P wave.

The Lagrangian for the $\rho NN$ vertex is \cite{Machleidt:1987hj}
\begin{equation}
\mathcal{L}_{\rho NN}=\sqrt{2}g_{\rho  NN}\bar n [\gamma _{\mu }-\frac{\kappa _{\rho }}{2m_{N}}i \sigma _{\mu \nu }q^{\nu }]p \rho ^{\mu },
\end{equation}
where $g^{2}_{\rho NN}/4\pi=0.92$ and $\kappa _{\rho }=6.1$ \cite{Guidal:1997hy,Machleidt:2000ge}. 

The Regge propagator for a spin-1 particle is slightly different from the scalar and pseudo-scalar one and is given by
\begin{equation}
\mathcal{P}^{\rho }_{Regge}=(\frac{s}{s_{0}})^{\alpha _{\rho } (t)-1}\frac{\pi \alpha '_{\rho }}{\sin (\pi (\alpha _{\rho } (t)-1))}\frac{\mathcal{S}_{\rho }+e^{-i\pi (\alpha _{\rho } (t)-1)}}{2\Gamma (\alpha _{\rho }(t))}. 
\end{equation}
Again the two trajectories with opposite signatures can be degenerate and form a common trajectory.
The $\rho $ Regge trajectory is $\alpha _{\rho }(t)=0.55+0.8\;t$ \cite{Guidal:1997hy}.

In the cross section calculation we also add the following form factors respectively to the $Z\rho \psi '$ and $\rho NN$ vertices:
\begin{equation}
\frac{M^{2}_{\psi }-m^{2}_{\rho }}{M^{2}_{\psi }-t}
\end{equation}
and
\begin{equation}
\frac{\Lambda ^{2}_{\rho }-m^{2}_{\rho }}{\Lambda ^{2}_{\rho }-t},
\end{equation}
where $\Lambda _{\rho }=1.31\;GeV$ \cite{Machleidt:2000ge}.

\subsubsection{\label{subsubsec:rho1-}The $J^{P}=1^{-}$ case.}
For the $Z(4430)$ with $J^{P}=1^{-}$ quantum numbers, we have to choose a vector-vector-vector interaction for the $Z\psi '\rho $ vertex \cite{Ma:2010xx} 
\begin{equation}
\mathcal{L}_{Z\psi '\rho  }=-i\frac{g_{Z\psi '\rho  }}{M_{Z}}[(\psi '^{\nu }\partial^{^{^{\hspace{-0.2cm}\leftrightarrow}}} _{\mu } Z_{\nu })\rho ^{\mu }+(\rho _{\nu }\partial^{^{^{\hspace{-0.2cm}\leftrightarrow}}} _{\mu } \psi '^{\nu })Z ^{\mu }+(Z^{\nu }\partial^{^{^{\hspace{-0.2cm}\leftrightarrow}}} _{\mu } \rho _{\nu })\psi '^{\mu }].
\end{equation}
The constant $g_{Z\psi' \rho }$ is unknown and in this case taken equal, as an upper limit, to the corresponding constant for the pion exchange $g_{Z\psi' \pi }/M_{Z}=2.365\;GeV^{-1}$.

\subsubsection{\label{subsubsec:rho1+}The $J^{P}=1^{+}$ case.}
If the $Z(4430)$ will be found to be an axial vector, the Lagrangian for the $Z\psi '\rho $ vertex would be \cite{Ma:2010xx}
\begin{equation}
\mathcal{L}_{Z\psi '\rho  }=\frac{g_{Z\psi '\rho  }}{M_{Z}}\epsilon _{\nu \mu \rho \alpha }Z^{\mu }\rho ^{\nu }(\partial ^{\alpha }\psi '^{\beta }-\partial ^{\beta }\psi '^{\alpha }).
\end{equation}
As done before, we use for the unknown constant $g_{Z\psi' \rho }$ the value of the constant $g_{Z\psi' \pi }/M_{Z}=2.01\;GeV^{-1}$ of the corresponding pionic exchange.

\subsubsection{\label{subsubsec:rho0-}The $J^{P}=0^{-}$ case.}
In the $J^{P}=0^{-}$ case, the Lagrangian for the $Z$ vertex is \cite{Guidal:1997hy}
\begin{equation}
\mathcal{L}_{Z\psi '\rho  }=\frac{g_{Z\psi' \rho }}{M_{Z}}\epsilon ^{\nu \mu \alpha \beta }\partial _{\mu }\psi '_{\nu }\partial _{\alpha }\rho _{\beta }Z,
\end{equation}
where $g_{Z\psi' \rho }$ is unknown and in this case taken equal to $g_{Z\psi' \pi }/M_{Z}=1.39\;GeV^{-1}$.

\begin{figure}
\includegraphics[width=10cm]{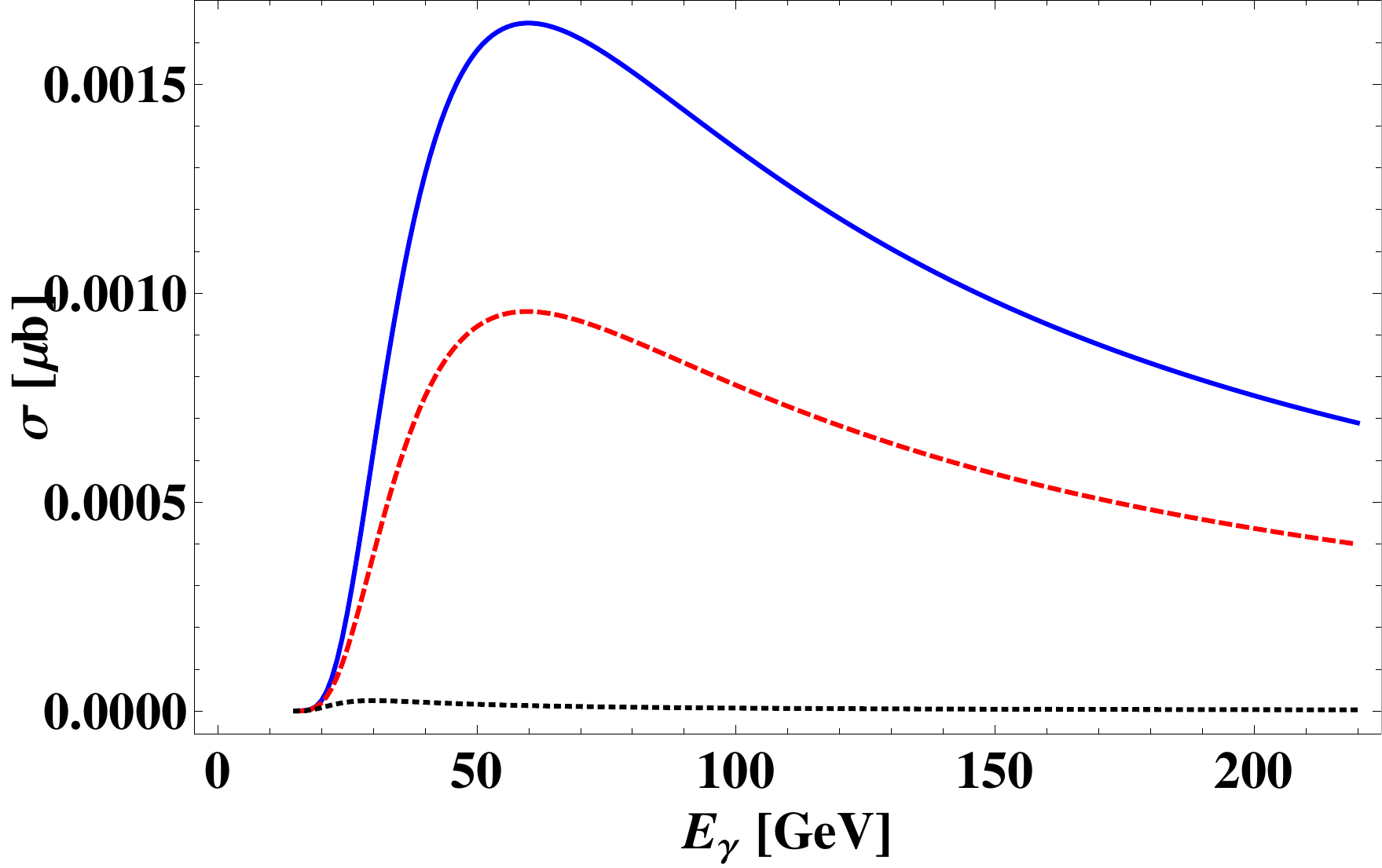}  
\caption{\label{fig:ZTotCrosssecsrho}Total cross section for the photoproduction of $Z^{+}(4430)$ with $J^{P}=1^{-}$ (Blue solid line), $J^{P}=1^{+}$ (Red dashed line) and $J^{P}=0^{-}$ (Black dotted line) through $\rho $ exchange.}
\end{figure}

The resulting total cross section in the three cases $J^{P}=1^{-},1^{+},0^{-}$ are given in Fig. \ref{fig:ZTotCrosssecsrho}.

\subsection{\label{subsec:risultaticrosssec}Numerical Results.}
Combining the $\pi ,\rho $ and $a_{0}$ Regge trajectories exchange terms for the photoproduction of the $Z^{+}(4430)$, we obtain the invariant matrix for the total process:
\begin{eqnarray}
\sum\limits_{pol}|\mathcal{M}_{\pi \rho a_{0}}|^{2} & = & \sum\limits_{pol}|\mathcal{M}_{\pi}+\mathcal{M}_{\rho}+\mathcal{M}_{a_{0}}|^{2}= \nonumber \\
& = & \sum\limits_{pol}(|\mathcal{M}_{\pi}|^{2}+|\mathcal{M}_{\rho}|^{2}+|\mathcal{M}_{a_{0}}|^{2}+2Re(e^{i(\alpha _{\pi }(t)-\alpha _{\rho }(t))}\mathcal{M}_{\pi}^{\dagger }\mathcal{M}_{\rho})+ \nonumber \\
& + & 2Re(e^{i(\alpha _{\pi }(t)-\alpha _{a_{0}}(t))}\mathcal{M}_{\pi}^{\dagger }\mathcal{M}_{a_{0}})+2Re(e^{i(\alpha _{a_{0}}(t)-\alpha _{\rho }(t))}\mathcal{M}_{a_{0}}^{\dagger }\mathcal{M}_{\rho})),
\end{eqnarray}  
where the $a_{0}$ channel and its interference terms are present only in the $J^{P}=1^{-},1^{+}$ cases.
We find that the pionic channel is by far predominant for every $J^{P}$ quantum numbers while the $\pi \rho $ and $\pi a_{0}$ interference terms are 0. 

\begin{figure}
\includegraphics[width=10cm]{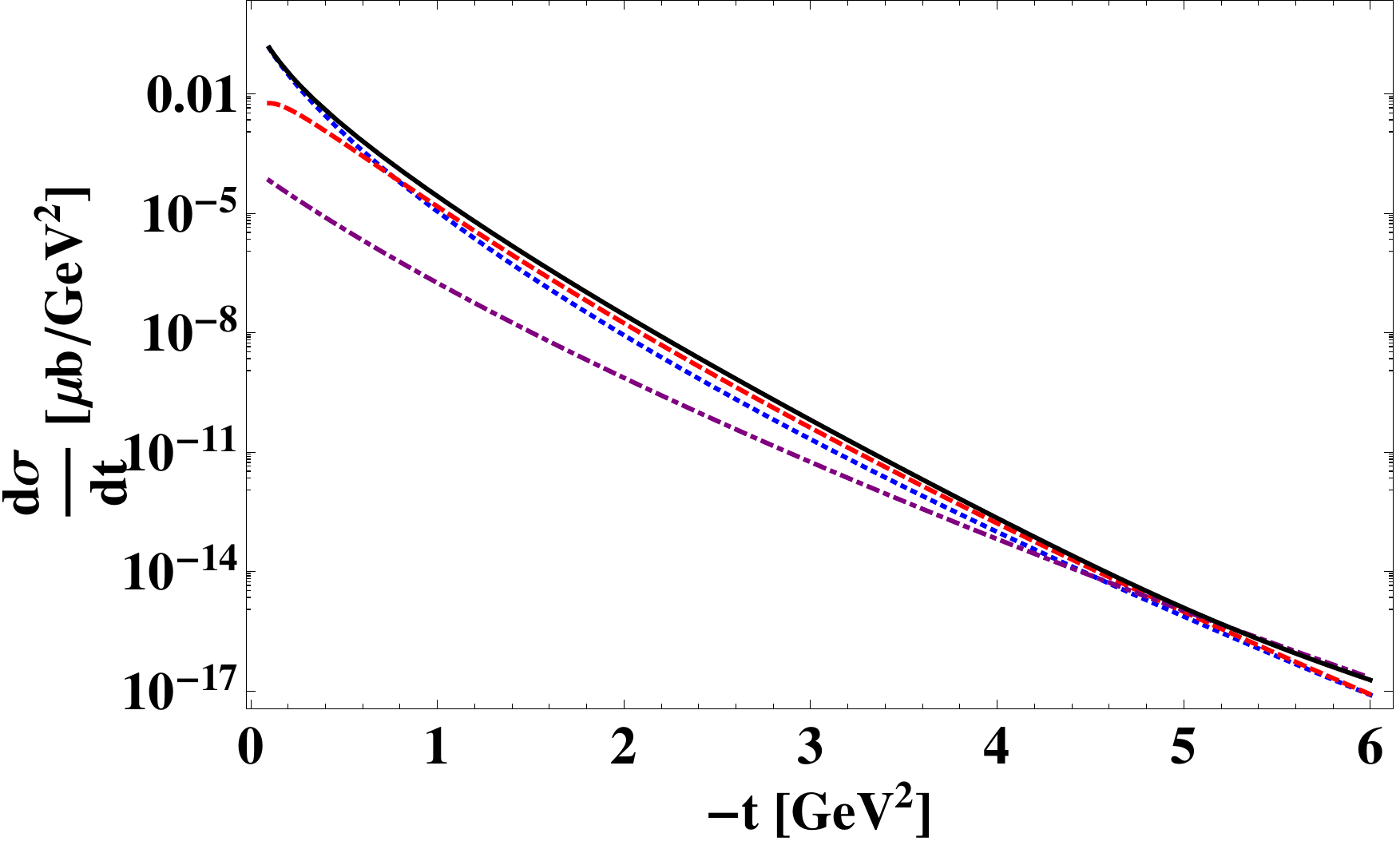}  
\caption{\label{fig:ZDiffAll1minus}Differential cross section for the photoproduction of $Z^{+}(4430)$ through the $\pi +\rho +a_{0}$ channels with $J^{P}=1^{-}$, at photon energy $E_{\gamma }=40\;GeV$. The blue dotted, red dashed and purple dot-dashed lines represent respectively the sole $\pi $, $\rho $ and $a_{0}$ exchange contributions, while the black solid line is the complete $\pi \rho a_{0}$ differential cross section.}
\end{figure}

\begin{figure}
\includegraphics[width=10cm]{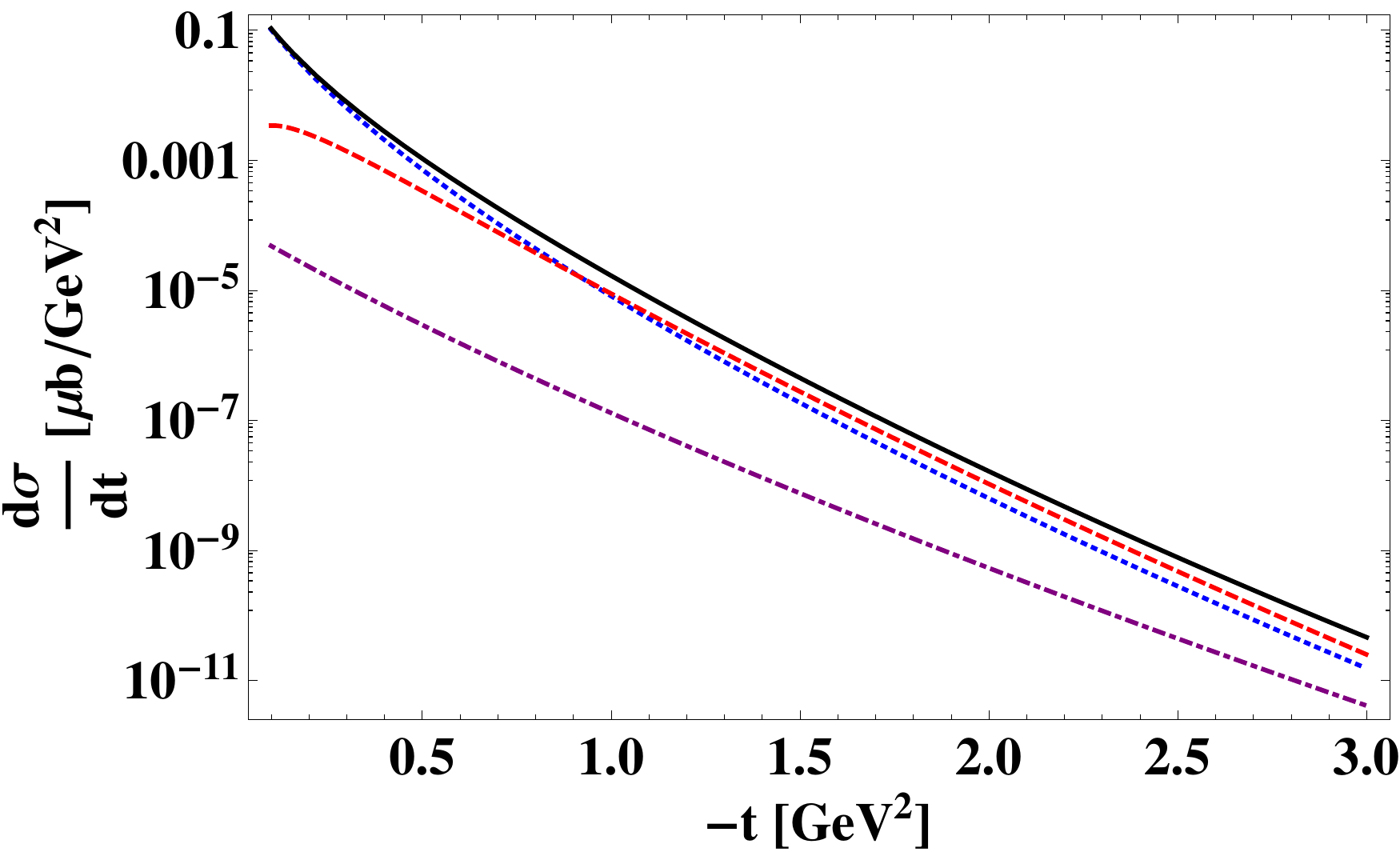}  
\caption{\label{fig:ZDiffAll1plus}Differential cross section for the photoproduction of $Z^{+}(4430)$ through the $\pi +\rho +a_{0}$ channels with $J^{P}=1^{+}$, at photon energy $E_{\gamma }=40\;GeV$. The blue dotted, red dashed and purple dot-dashed lines represent respectively the sole $\pi $, $\rho $ and $a_{0}$ exchange contributions, while the black solid line is the complete $\pi \rho a_{0}$ differential cross section.}
\end{figure}

\begin{figure}
\includegraphics[width=10cm]{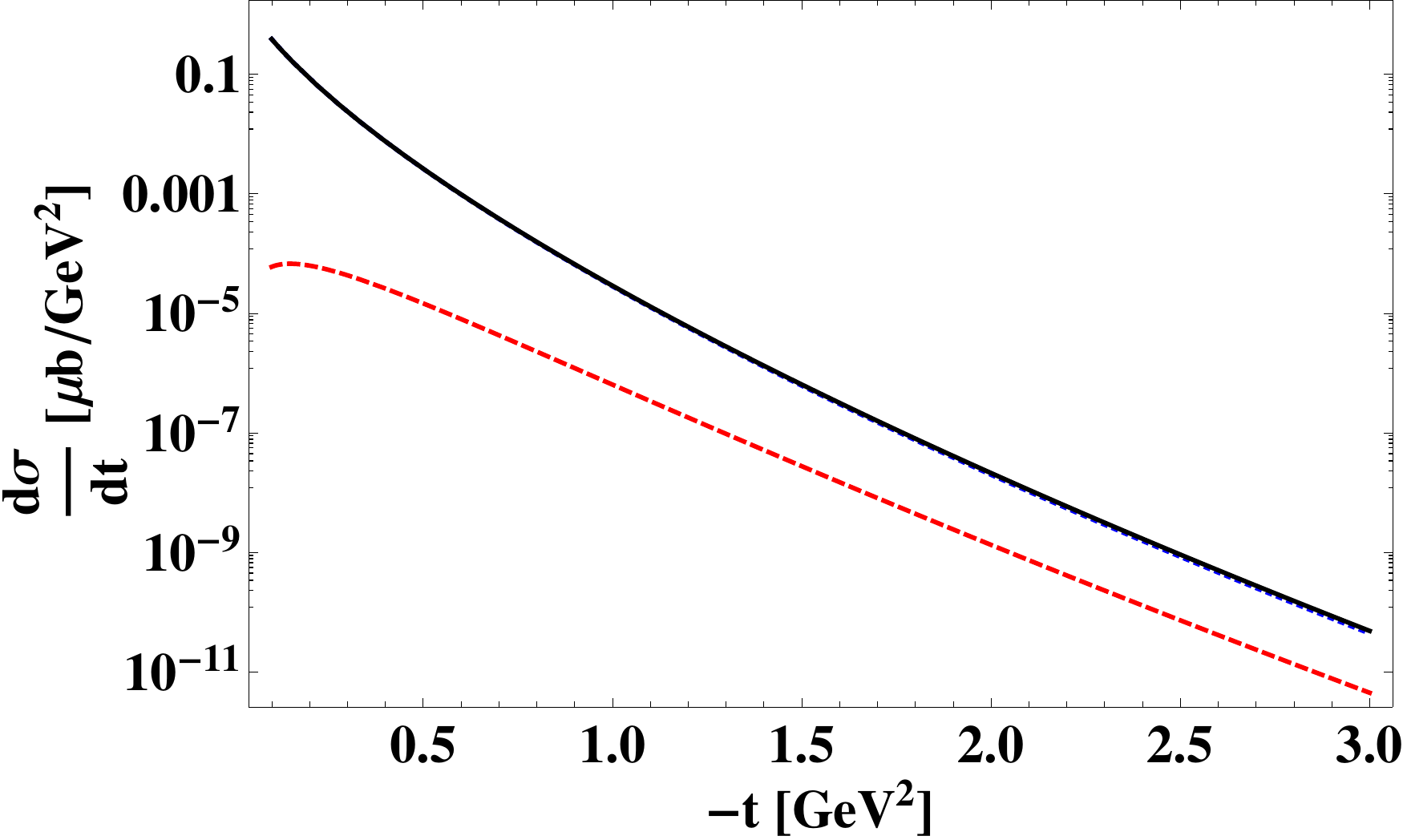}  
\caption{\label{fig:ZDiffAll0minus}Differential cross section for the photoproduction of $Z^{+}(4430)$ through the $\pi +\rho $ channels with $J^{P}=0^{-}$, at photon energy $E_{\gamma }=40\;GeV$. The blue dotted and red dashed lines represent respectively the sole $\pi $ and $\rho $ exchange contributions, while the black solid line is the complete $\pi \rho$ differential cross section.}
\end{figure}

\begin{figure}
\includegraphics[width=10cm]{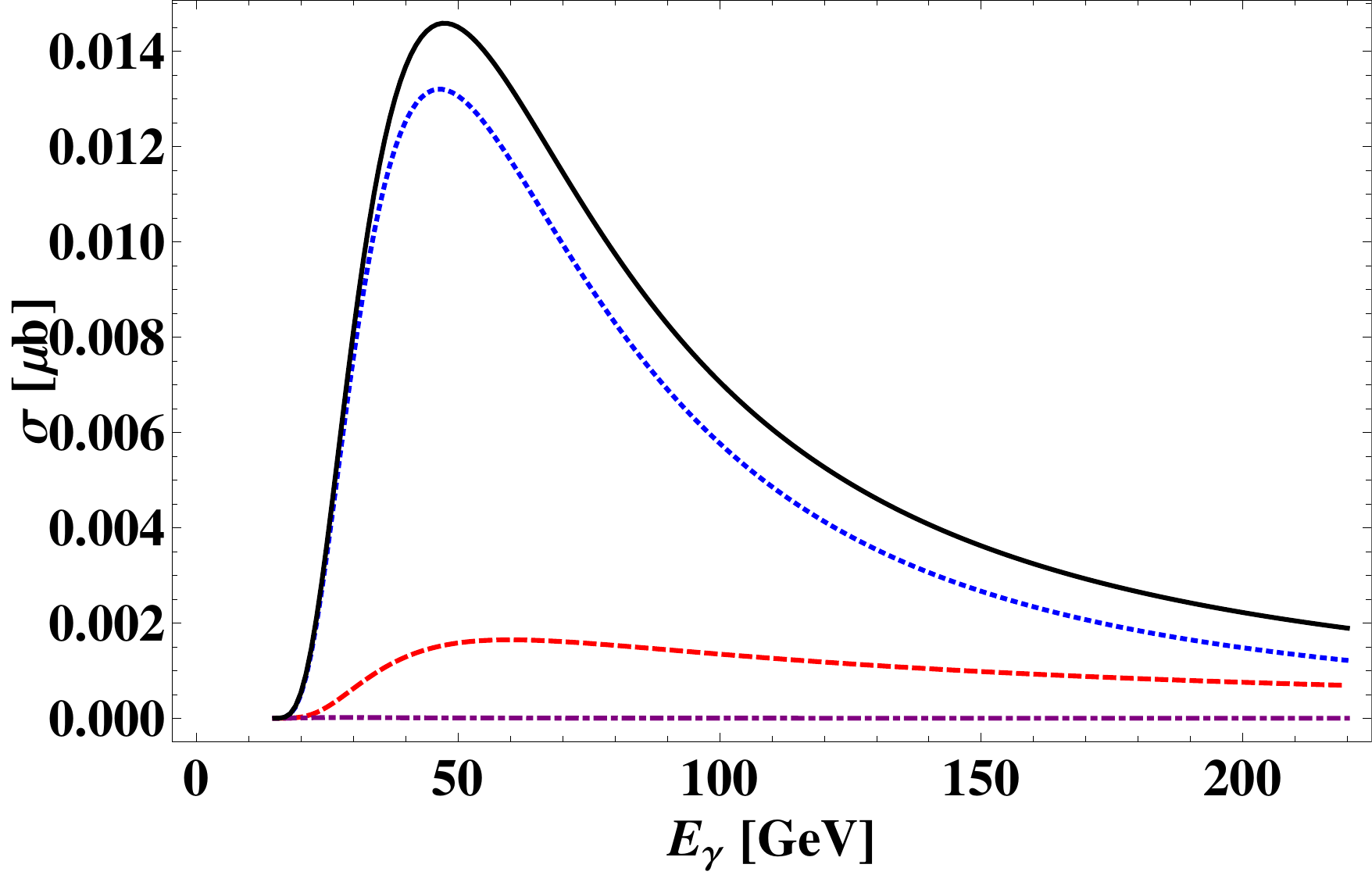}  
\caption{\label{fig:ZTotAll1minus}Total cross section for the photoproduction of $Z^{+}(4430)$ through the $\pi +\rho +a_{0}$ channels with $J^{P}=1^{-}$, at photon energy $E_{\gamma }=40\;GeV$. The blue dotted, red dashed and purple dot-dashed lines represent respectively the sole $\pi $, $\rho $ and $a_{0}$ exchange contributions, while the black solid line is the complete $\pi \rho a_{0}$ total cross section.}
\end{figure}

\begin{figure}
\includegraphics[width=10cm]{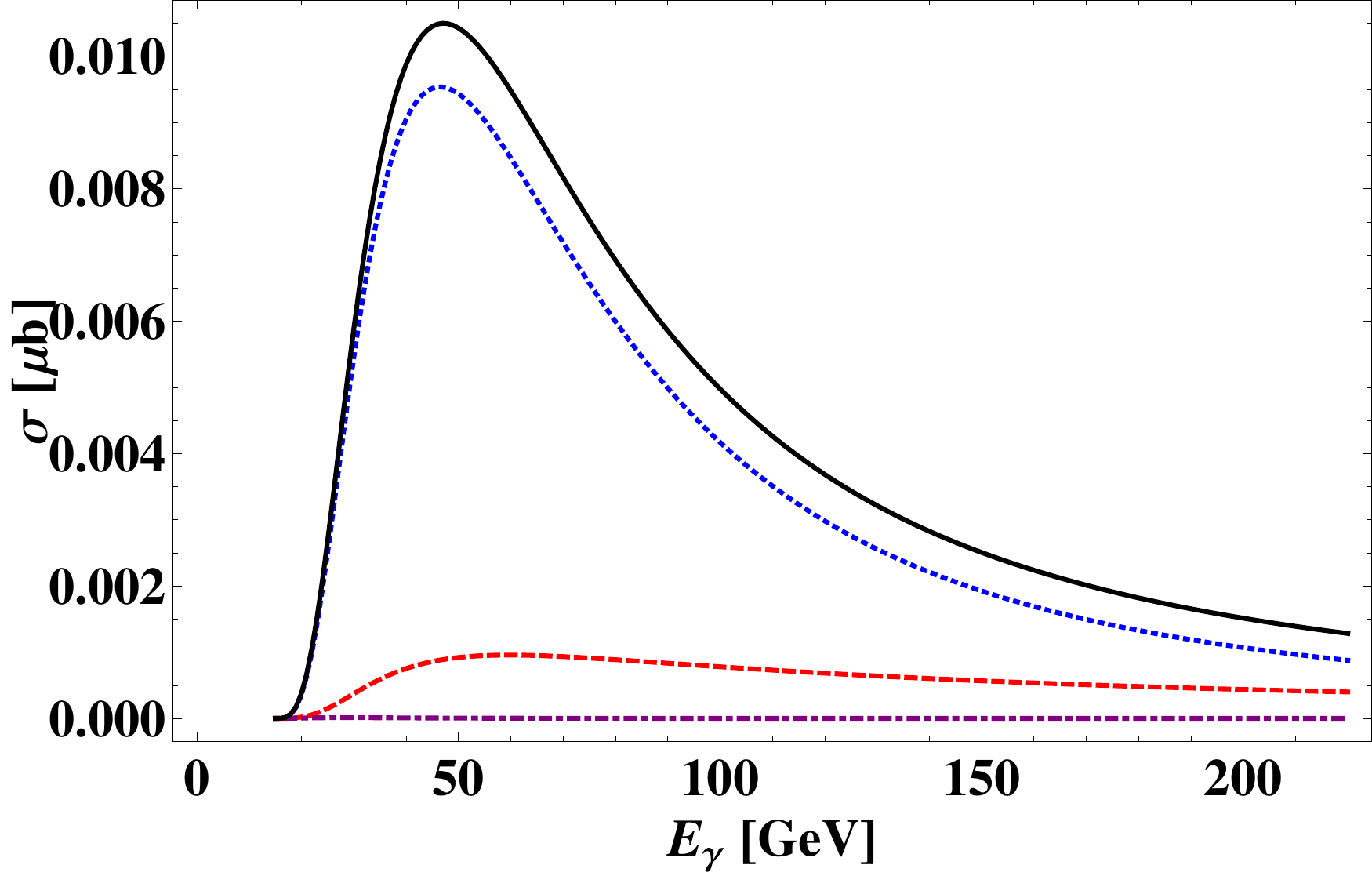}  
\caption{\label{fig:ZTotAll1plus}Total cross section for the photoproduction of $Z^{+}(4430)$ through the $\pi +\rho +a_{0}$ channels with $J^{P}=1^{+}$, at photon energy $E_{\gamma }=40\;GeV$. The blue dotted, red dashed and purple dot-dashed lines represent respectively the sole $\pi $, $\rho $ and $a_{0}$ exchange contributions, while the black solid line is the complete $\pi \rho a_{0}$ total cross section.}
\end{figure}

\begin{figure}
\includegraphics[width=10cm]{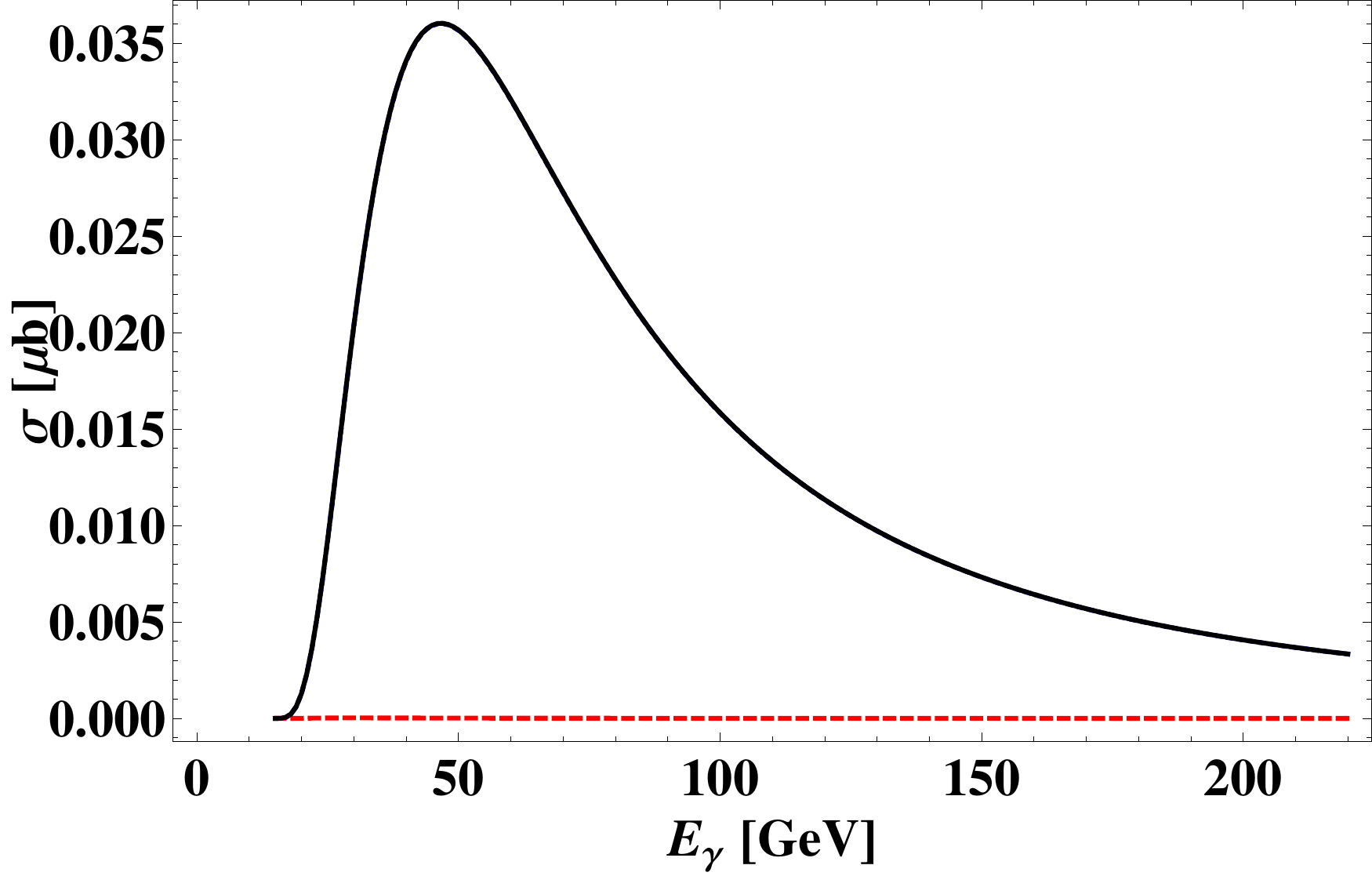}  
\caption{\label{fig:ZTotAll0minus}Total cross section for the photoproduction of $Z^{+}(4430)$ through the $\pi +\rho $ channels with $J^{P}=0^{-}$, at photon energy $E_{\gamma }=40;GeV$. The blue dotted and red dashed lines represent respectively the sole $\pi $ and $\rho $ exchange contributions, while the black line is the complete $\pi \rho$ total cross section.}
\end{figure}

The results for the differential cross sections for the photoproduction of the $Z^{+}(4430)$ are shown in Figs. \ref{fig:ZDiffAll1minus}, \ref{fig:ZDiffAll1plus} and \ref{fig:ZDiffAll0minus}, respectively in the $1^{-}$, $1^{+}$ and $0^{-}$ cases. The total cross sections are in Figs. \ref{fig:ZTotAll1minus}, \ref{fig:ZTotAll1plus} and \ref{fig:ZTotAll0minus}.

% \begin{figure}
% \includegraphics[width=10cm]{ZTotAll1minusLog.pdf}  
% \caption{\label{fig:ZTotAll1minusLog}Total cross section for the photoproduction of $Z^{+}(4430)$ through the $\pi +\rho +a_{0}$ channels with $J^{P}=1^{-}$, at photon energy $E_{\gamma }=40\;GeV$. The blue, red and purple lines represent respectively the sole $\pi $, $\rho $ and $a_{0}$ exchange contributions, while the black line is the complete $\pi \rho a_{0}$ total cross section. (ALTERNATIVE FIGURE WITH A LOGARITHMIC SCALE)}
% \end{figure}

% \begin{figure}
% \includegraphics[width=10cm]{ZTotAll1plusLog.pdf}  
% \caption{\label{fig:ZTotAll1plusLog}Total cross section for the photoproduction of $Z^{+}(4430)$ through the $\pi +\rho +a_{0}$ channels with $J^{P}=1^{+}$, at photon energy $E_{\gamma }=40\;GeV$. The blue, red and purple lines represent respectively the sole $\pi $, $\rho $ and $a_{0}$ exchange contributions, while the black line is the complete $\pi \rho a_{0}$ total cross section. (ALTERNATIVE FIGURE WITH A LOGARITHMIC SCALE)}
% \end{figure}

% \begin{figure}
% \includegraphics[width=10cm]{ZTotAll0minusLog.pdf}  
% \caption{\label{fig:ZTotAll0minusLog}Total cross section for the photoproduction of $Z^{+}(4430)$ through the $\pi +\rho $ channels with $J^{P}=0^{-}$, at photon energy $E_{\gamma }=40\;GeV$. The blue and red lines represent respectively the sole $\pi $ and $\rho $ exchange contributions, while the black line is the complete $\pi \rho$ total cross section. (ALTERNATIVE FIGURE WITH A LOGARITHMIC SCALE)}
% \end{figure}

\section{\label{sec:asimmetrie}Polarized cross sections and asymmetries.}
The presence of the $\rho $ and $a_{0}$ terms becomes determinant in the calculation of the polarization asymmetries. In fact the $\pi $ channel contribution is a constant (more often 0), thus the behavior of the asymmetries is determined by the other components of the cross section.
\subsection{\label{subsec:asimmetriafot}The photon asymmetry.}
The formula for the photon asymmetry is
\begin{equation}
\Sigma =\frac{1}{P_{\gamma }}\frac{\sigma _{\perp }-\sigma _{\parallel }}{\sigma _{\perp }+\sigma _{\parallel }},
\end{equation}
where the polarized cross sections are calculated respectively normal and parallel to the production plane, defined by $\vec k_{1}$ and $\vec k_{2}$.
We consider a full polarization of the beam, thus the polarization degree is $P_{\gamma }=1$. 

\begin{figure}
\includegraphics[width=10cm]{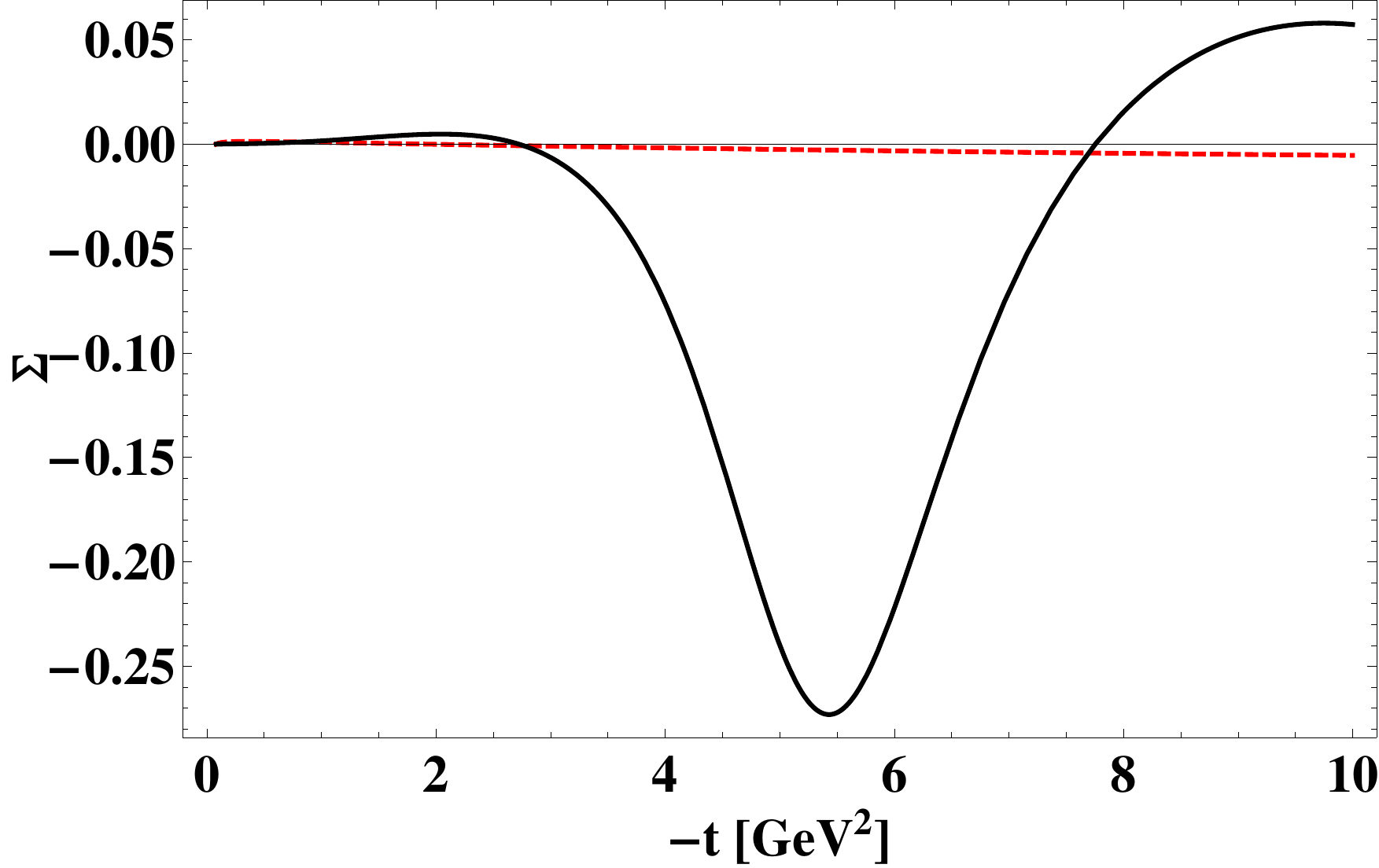}  
\caption{\label{fig:Assfot1minus}Photon asymmetry for the photoproduction of $Z^{+}(4430)$ with $J^{P}=1^{-}$ at photon energy $40\;GeV$: in solid black the $\pi +\rho +a_{0}$ exchange, in dashed red the $\rho $ contribution.}
\end{figure}

\begin{figure}
\includegraphics[width=10cm]{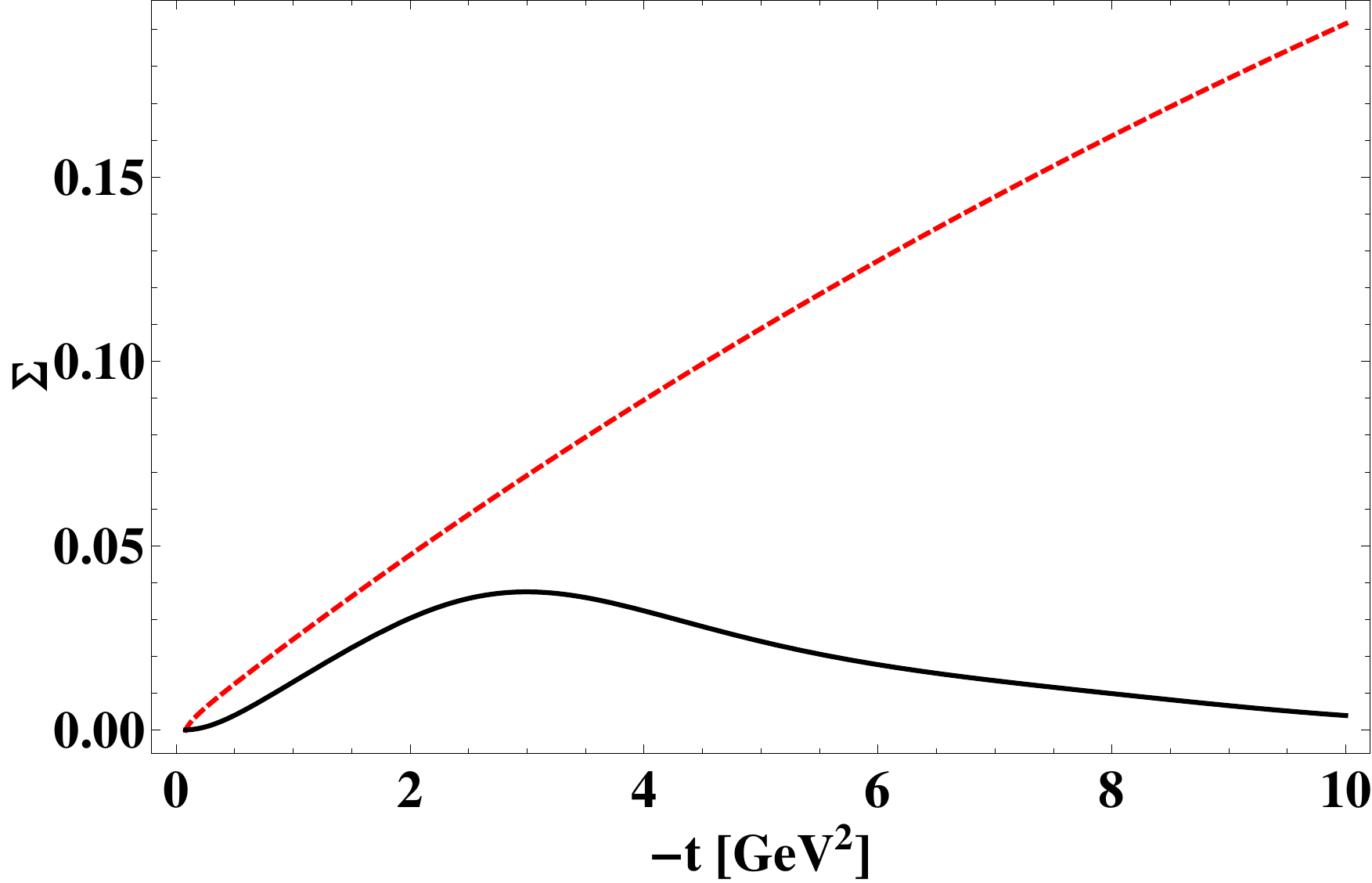}  
\caption{\label{fig:Assfot1plus}Photon asymmetry for the photoproduction of $Z^{+}(4430)$ with $J^{P}=1^{+}$ at photon energy $40\;GeV$: in solid black the $\pi +\rho +a_{0}$ exchange, in dashed red the $\rho $ contribution.}
\end{figure}

The $J^{P}=1^{+}, 1^{-}$ polarized cross sections with pion and $a_{0}$ exchange are independent from the polarization state of the photon, thus in these cases the asymmetry is 0. The only contributions to the asymmetry come from the $\rho $ trajectory and the interference terms. In Figs. \ref{fig:Assfot1minus} and \ref{fig:Assfot1plus} we show respectively the $1^{-}$ and $1^{+}$ photon asymmetry. 

\begin{figure}
\includegraphics[width=10cm]{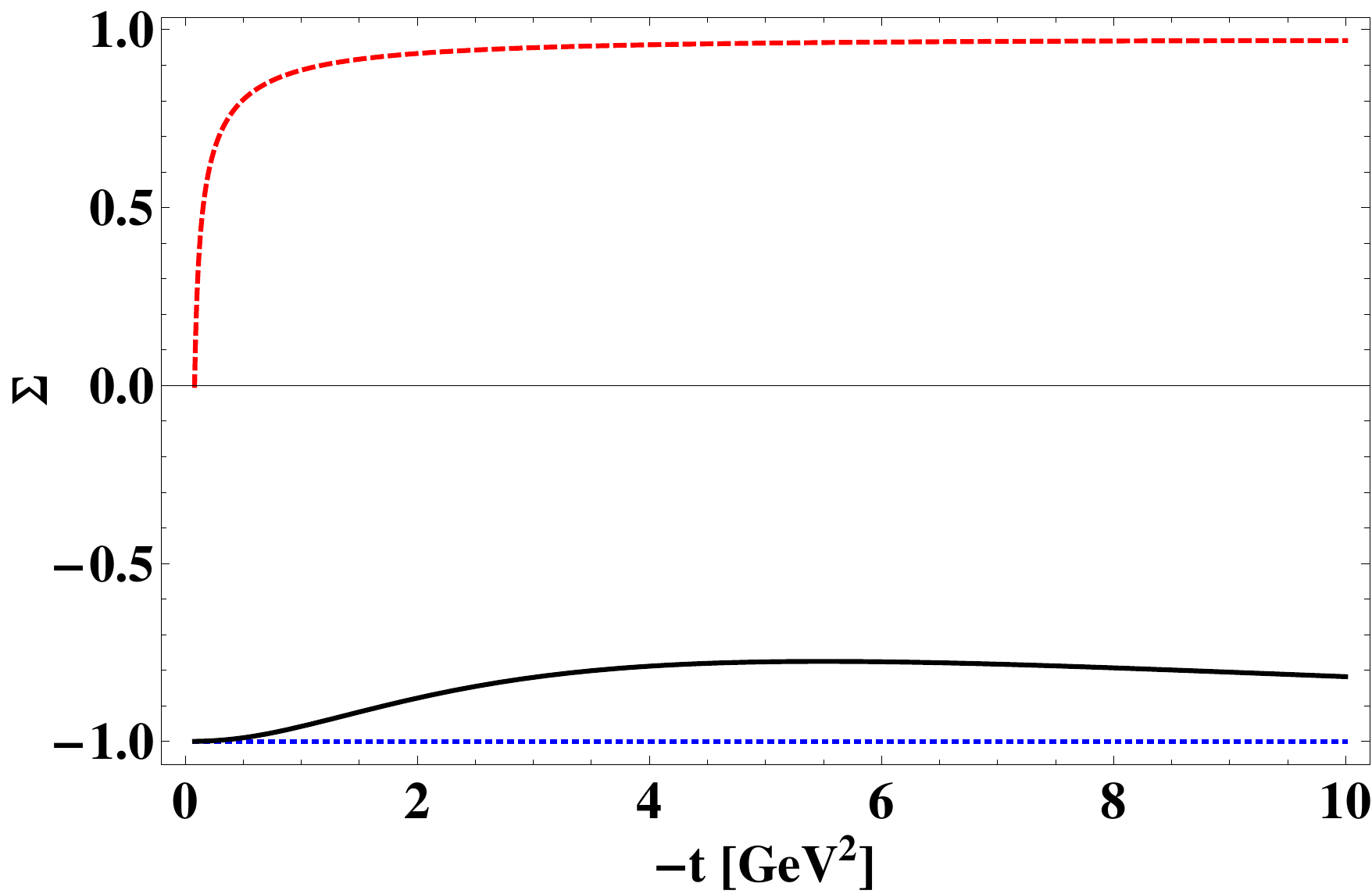}  
\caption{\label{fig:Assfot0minus}Photon asymmetry for the photoproduction of $Z^{+}(4430)$ with $J^{P}=0^{-}$ at photon energy $40\;GeV$: in solid black the $\pi +\rho$ exchange, in dotted blue and dashed red respectively the $\pi $ and $\rho $ contributions.}
\end{figure}

In the $J^{P}=0^{-}$ case both the $\pi $, whose asymmetry is a constant $\Sigma =-1$, and the $\rho $ contribute to the total $\pi +\rho$ photon asymmetry. The results can be seen in Fig. \ref{fig:Assfot0minus}.

\subsection{\label{subsec:asimmetriaTpol}The target and recoil asymmetries.}
\begin{figure}
\includegraphics[width=10cm]{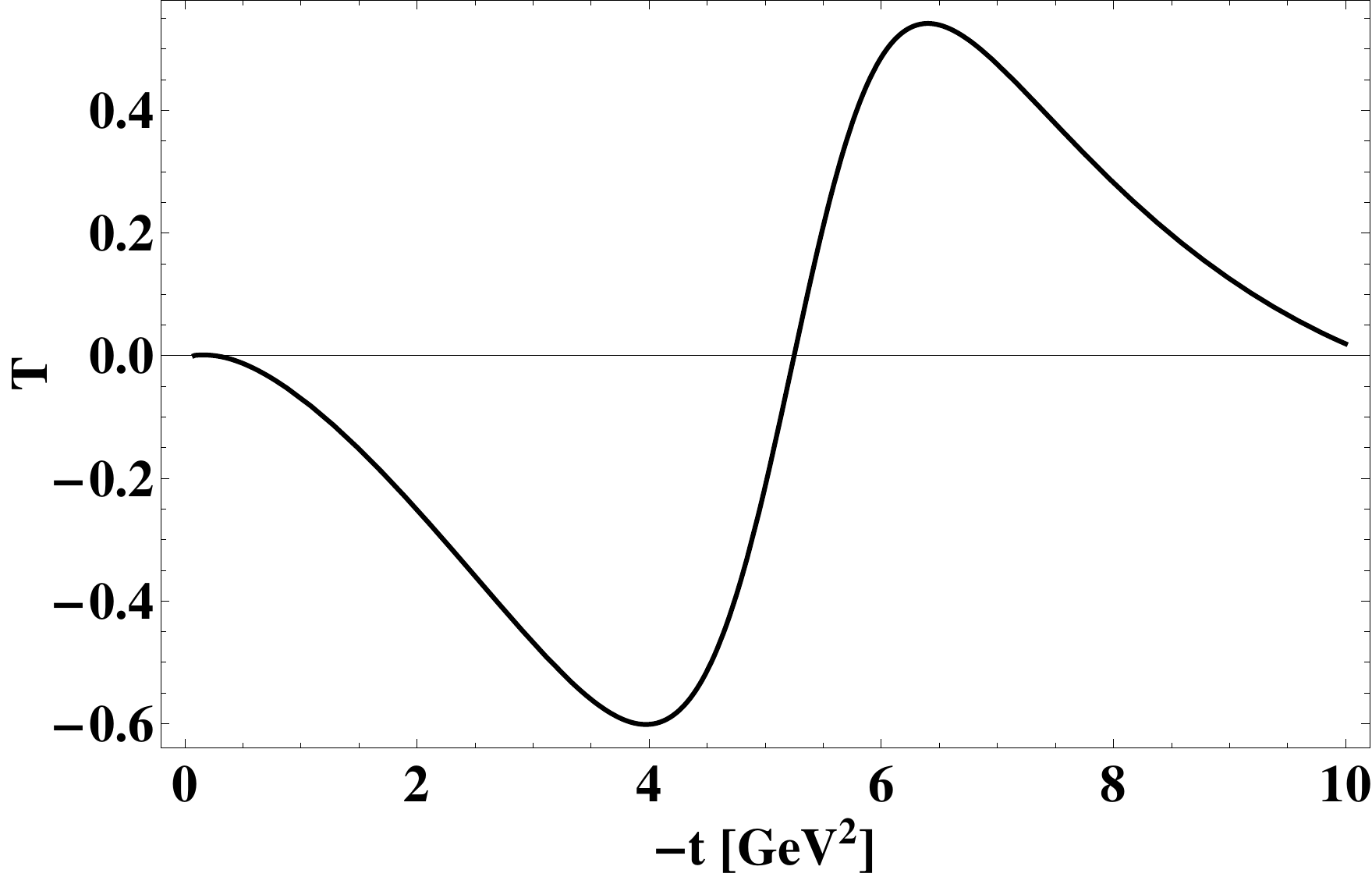}  
\caption{\label{fig:AssTpol1minus}Target asymmetry for the photoproduction of $Z^{+}(4430)$ with $J^{P}=1^{-}$ at photon energy $40\;GeV$.}
\end{figure}

\begin{figure}
\includegraphics[width=10cm]{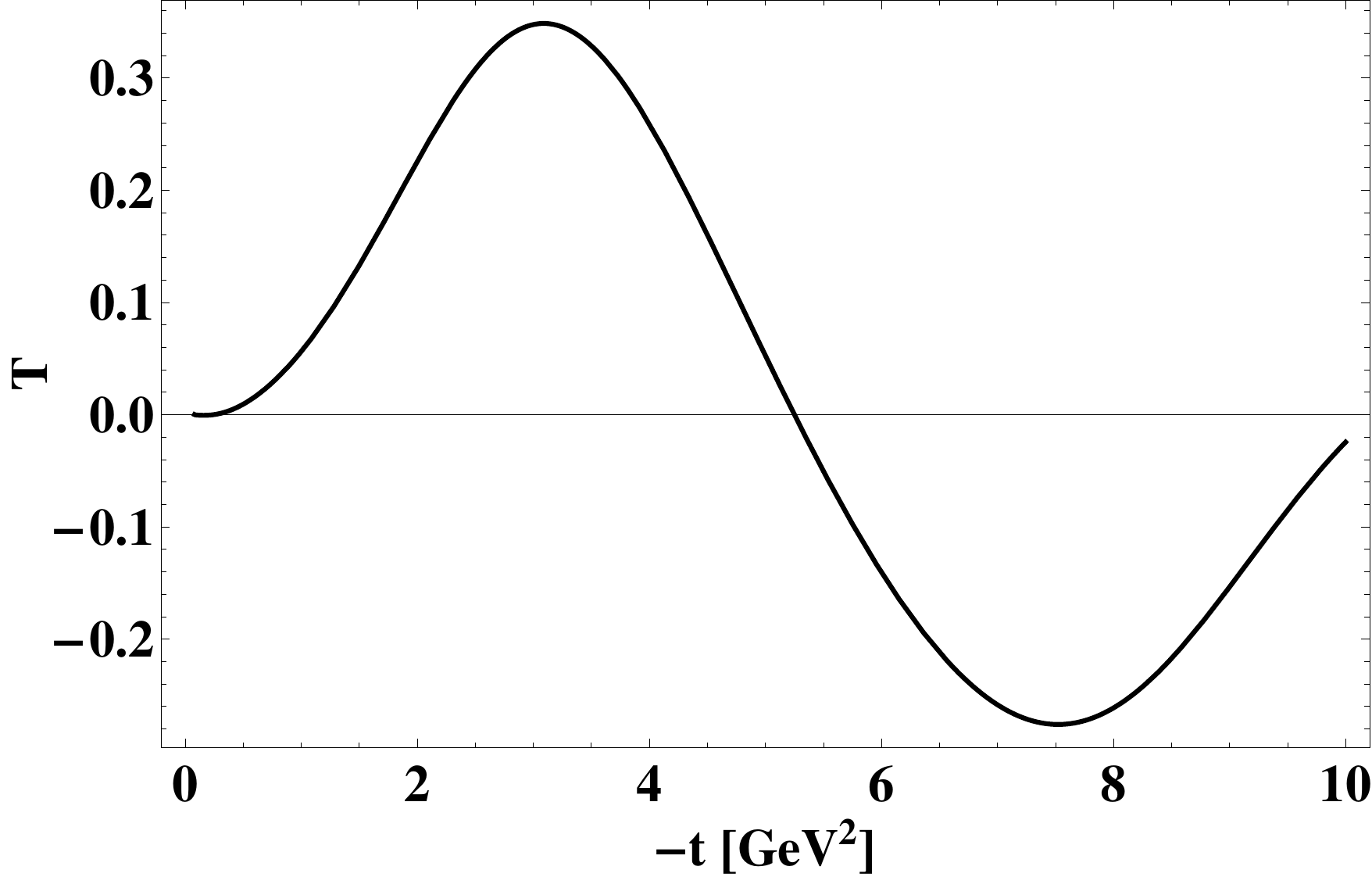}  
\caption{\label{fig:AssTpol1plus}Target asymmetry for the photoproduction of $Z^{+}(4430)$ with $J^{P}=1^{+}$ at photon energy $40\;GeV$.}
\end{figure}

The target asymmetry is defined as
\begin{equation}
T =\frac{\sigma _{\uparrow }-\sigma _{\downarrow }}{\sigma _{\uparrow }+\sigma _{\downarrow }},
\end{equation}
where the up and down target nucleon spins are calculated parallel and antiparallel to the vector $\vec k_{1}\times \vec k_{2}$.
It is well known that only the interference between amplitudes produce a target asymmetry. For the $Z^{+}(4430)$ photoproduction the only non-zero term is the $\rho -a_{0}$ interference term, thus the target asymmetry for $J^{P}=0^{-}$ is zero. The target asymmetry for $J^{P}=1^{-}$ and $1^{+}$ is shown in Figs. \ref{fig:AssTpol1minus} and \ref{fig:AssTpol1plus}.

The recoil asymmetry is defined as
\begin{equation}
P =\frac{\sigma _{\uparrow }-\sigma _{\downarrow }}{\sigma _{\uparrow }+\sigma _{\downarrow }},
\end{equation}
where the spin of the recoil nucleon is again projected onto the vector $\vec k_{1}\times \vec k_{2}$. For all $J^{P}$ quantum numbers the recoil asymmetry results are equal to the target asymmetry. This is what should actually be expected since the difference between the target and the recoil asymmetries has been proved to depend only on the exchange of $1^{++}$ (i.e. $a_{1}(1260)$) and $2^{--}$ mesons \cite{Sibirtsev:2007wk}.

\section{\label{sec:conclusione} Conclusion.}
In this work we have calculated the differential and total cross sections for the photoproduction of the resonance $Z^{+}(4430)$ in the reaction $\gamma p\rightarrow Z^{+}(4430)n$. We have used an effective Lagrangian approach modified by the exchange of Regge trajectories, instead of singular particles. Thus the usual Feynman propagator had to be replaced with a Regge propagator, which economically includes the effects of the exchange of the high spin and mass particles belonging to the Regge trajectory. These calculations have been accomplished for each possible $J^{P}$ quantum number of the $Z(4430)$ in $S$ and $P$ wave, i.e. $J^{P}=1^{-},1^{+},0^{-},2^{-}$ in the case of pion exchange, and combining the exchanges of $\pi $, $\rho $ and $a_{0}(980)$ trajectories. Comparing the cross sections obtained in the present work with the corresponding ones calculated by Liu, Zhao and Close \cite{Liu:2008qx}, we can note that the effect of introducing the Regge trajectories has been to sensibly reduce the magnitude of the cross sections. The Regge method used in this work has already been proved more precise than the standard effective Lagrangian approach in the pion and kaon photoproduction \cite{Guidal:1997hy}, then the observation or not of the $Z(4430)$ with the cross sections predicted here could be a further confirmation or a denial of the validity of this approximate method. 
The polarization observables, such as the photon, target and recoil asymmetries, have also been obtained from the cross sections, in the same cases specified above, with the exception of the $2^{-}$ whose cross section is unimportant. 

The photoproduction of $Z(4430)$ begins at a photon energy in the laboratory of circa $15\;GeV$ and has the maximum around $30-40\;GeV$. In the current situation no experimental facility can produce real photons at such high energy. However these energy ranges were possible at HERA, which could produce high energy quasi-real photons with low virtuality \cite{:2011pz}. Thus new analysis of the data in the channels studied in this work could be interesting. After the shutdown of HERA, new data for high energy photoproduction of vector mesons will probably be produced at LHC \cite{Nystrand:2005gv,Baltz:2007kq}.  

With this Regge-based model we can get good results for the amplitudes at high energies and forward angles without using any free parameter. All constants are actually fixed to independent experimental observables. The only exceptions are the coupling constants of the vertices $Z\psi '\rho $ and $Z\psi 'a_{0} $, for which we do not have any experimental data, that have been set as an upper limit equal to the constant for the $Z\psi '\pi $.

As it could be expected, we find the $\pi$ Regge trajectory exchange channel by far dominating in the cross section, even considering that the $\rho $ and $a_{0}$ contributions are overestimated due to the choice of their coupling constants. The interference terms with the pion channel are found to be zero. On the contrary the $\rho $ and $a_{0}$ exchanges, and their interference terms, become decisive in the determination of the polarization observables.     

\bibliography{Bibliografia}

\printfigures*
\end{document}